\documentclass[twocolumn, twocolappendix]{aastex631}

\usepackage{amsmath}
\usepackage{graphicx}
\usepackage{orcidlink}

\let\tablenum\relax 
\usepackage[separate-uncertainty=true]{siunitx}
\DeclareSIUnit{\Msun}{\text{\ensuremath{M_\odot}}}

\newcommand\ktuc{$\kappa$~Tuc} 

\newcommand{\sbs}[1]{\ensuremath{_\text{#1}}}

\newcommand{\prior}[2]{\ensuremath{\left[#1,#2 \right]}}

\newcommand{\snm}[1]{\tablenotemark{\scriptsize #1}}

\newcommand{\phsnm}[1]{\phantom{\tablenotemark{\scriptsize #1}}}

\begin{document}

\title{
    Interferometric Detection and Orbit Modeling of the Subcomponent in the
    Hot-dust System $\kappa$ Tuc A:\\A Low-mass Star on an Eccentric Orbit in a
    Hierarchical-quintuple System
}

\correspondingauthor{T. A. Stuber}
\email{tstuber@arizona.edu}

\author[0000-0003-2185-0525]{T. A. Stuber}
\affiliation{
    Department of Astronomy and Steward Observatory,
    The University of Arizona,
    933 North Cherry Ave,
    Tucson, AZ 85721, USA
}

\author[0000-0003-2125-0183]{A. Mérand}
\affiliation{
    European Southern Observatory Headquarters,
    Karl-Schwarzchild-Str. 2,
    85748 Garching, Germany
}

\author[0000-0002-3036-0184]{F. Kirchschlager}
\affiliation{
    Sterrenkundig Observatorium,
    Ghent University,
    Krijgslaan 281-S9,
    9000 Gent, Belgium
}

\author[0000-0001-7841-3452]{S. Wolf}
\affiliation{
    Institute of Theoretical Physics and Astrophysics,
    Kiel University,
    Leibnizstr. 15,
    24118 Kiel, Germany
}

\author[0000-0001-8009-8383]{G. Weible}
\affiliation{
    Department of Astronomy and Steward Observatory,
    The University of Arizona,
    933 North Cherry Ave,
    Tucson, AZ 85721, USA
}

\author[0000-0002-4006-6237]{O. Absil}
\affiliation{
    STAR Institute,
    Universit\'e de Li\`ege,
    All\'ee du Six Ao\^ut 19c,
    4000 Li\`ege, Belgium
}

\author[0000-0001-5653-5635]{T. D. Pearce}
\affiliation{
    Department of Physics,
    University of Warwick,
    Gibbet Hill Road,
    Coventry CV4 7AL, UK
}

\author[0000-0001-6282-1339]{G. Garreau}
\affiliation{
    Institute of Astronomy,
    KU Leuven,
    Celestijnenlaan 200D,
    3001 Leuven, Belgium
}

\author[0000-0002-2725-6415]{J.-C. Augereau}
\affiliation{
    Univ. Grenoble Alpes, CNRS, IPAG,
    F-38000 Grenoble, France
}

\author[0000-0002-9209-5830]{W. C. Danchi}
\affiliation{
    NASA Goddard Space Flight Center,
    8800 Greenbelt Road,
    Greenbelt, MD 20771-2400, USA
}

\author[0000-0003-3499-2506]{D. Defrère}
\affiliation{
    Institute of Astronomy,
    KU Leuven,
    Celestijnenlaan 200D,
    3001 Leuven, Belgium
}

\author[0000-0001-6403-841X]{V. Faramaz-Gorka}
\affiliation{
    Department of Astronomy and Steward Observatory,
    The University of Arizona,
    933 North Cherry Ave,
    Tucson, AZ 85721, USA
}

\author[0000-0002-1272-6322]{J. W. Isbell}
\affiliation{
    Department of Astronomy and Steward Observatory,
    The University of Arizona,
    933 North Cherry Ave,
    Tucson, AZ 85721, USA
}

\author[0000-0002-3741-5950]{J. Kobus}
\affiliation{
    Institute of Theoretical Physics and Astrophysics,
    Kiel University,
    Leibnizstr. 15,
    24118 Kiel, Germany
}

\author[0009-0009-4573-2612]{A. V. Krivov}
\affiliation{%
    \mbox{
        Astrophysikalisches Institut und Universitätssternwarte,
        Friedrich-Schiller-Universität Jena,
        Schillergässchen 2–3,
        07745 Jena, Germany
    }
}

\author[0000-0002-2215-9413]{R. Laugier}
\affiliation{
    Institute of Astronomy,
    KU Leuven,
    Celestijnenlaan 200D,
    3001 Leuven, Belgium
}

\author[0009-0003-6954-5252]{K. Ollmann}
\affiliation{
    Institute of Theoretical Physics and Astrophysics,
    Kiel University,
    Leibnizstr. 15,
    24118 Kiel, Germany
}

\author[0000-0003-4759-6051]{R. G. Petrov}
\affiliation{
    Université Côte d'Azur,
    Observatoire de la Côte d'Azur,
    CNRS, Laboratoire Lagrange,
    Nice, France
}

\author[0009-0000-4227-5449]{P. Priolet}
\affiliation{
    Univ. Grenoble Alpes, CNRS, IPAG,
    F-38000 Grenoble, France
}

\author[0009-0000-3882-9242]{J. P. Scott}
\affiliation{
    Department of Astronomy and Steward Observatory,
    The University of Arizona,
    933 North Cherry Ave,
    Tucson, AZ 85721, USA
}

\author[0009-0002-9371-0740]{K. Tsishchankava}
\affiliation{
    Institute of Theoretical Physics and Astrophysics,
    Kiel University,
    Leibnizstr. 15,
    24118 Kiel, Germany
}

\author[0000-0002-2314-7289]{S. Ertel}
\affiliation{
    Department of Astronomy and Steward Observatory,
    The University of Arizona,
    933 North Cherry Ave,
    Tucson, AZ 85721, USA
}

\affiliation{
    Large Binocular Telescope Observatory,
    The University of Arizona,
    933 North Cherry Ave,
    Tucson, AZ 85721, USA
}


\begin{abstract}
The system \ktuc~A is part of a hierarchical-quintuple system and is a prime
target for studies of hot-exozodiacal dust, because a time-variable
near-infrared excess has been detected. We observed the system with the Multi
Aperture mid-Infrared Spectroscopic Experiment (MATISSE) and GRAVITY at the
Very Large Telescope Interferometer, and detected the stellar companion to the
primary \ktuc~Aa that was previously inferred by astrometry, \ktuc~Ab. Its
$L$-band flux ratio to the primary is \qty{1.32}{\percent} and its signature in
the MATISSE closure phases is mostly smaller than $\pm \qty{2}{\degree}$, which
makes \ktuc~Ab the highest-contrast companion ever detected with MATISSE
closure phases. We verified with GRAVITY that relative astrometry with
milliarcsecond precision can be retrieved from MATISSE closure phases.

Using multiple epochs of observations, we obtain a full orbital solution for
\ktuc~Ab. Its orbit has an eccentricity of $\num{0.94}$ and a semi-major axis
of \qty{4.8}{au}. The orbit of \ktuc~Ab and the orbit of the wider separation
companion \ktuc~B are mutually inclined. Based on the measured flux ratio of
\ktuc~Ab to Aa and their dynamical mass, we estimate the spectral type of
\ktuc~Ab to be M3.5$\,$V to M4.5$\,$V.

While the then unknown star \ktuc~Ab might have caused the putative detection
of hot-exozodiacal dust around \ktuc~Aa in 2012 and 2014, this cannot be for
the detection in 2019, giving rise to an intriguing system architecture. This
motivates studies investigating the interplay of the low-mass star on an
eccentric orbit, the hot-exozodiacal dust, and a possible planetesimal
reservoir.
\end{abstract}


\section{Introduction}
\label{sect_introduction}

Stellar surveys using near-infrared interferometry have repeatedly found
signatures in interferometric observables \citep{absil:2013, ertel:2014,
ertel:2016, nunez:2017, absil:2021} that are, since their first detection by
\citet{absil:2006}, commonly attributed to the presence of hot-exozodiacal dust
(a \textit{hot exozodi}) in the close vicinity of the star (see for reviews
\citeauthor{kral:2017} \citeyear{kral:2017} and \citeauthor{ertel:2025}
\citeyear{ertel:2025}). Since stellar companions to the host star can cause a
similar signature, they can be a source of confusion
\citep[e.g.,][]{tsishchankava:2025}. Thus, stars with known companions are
usually excluded from surveys searching for hot exozodis a priori
\citep[e.g.,][]{absil:2013, ertel:2014} and the obtained data are checked for
signs of new companions \citep[e.g.,][]{marion:2014}.

However, stars might host unknown companions, as those can avoid
interferometric detection for several reasons. The strength of the companion
signature in interferometric measurements generally depends on the telescope
positions projected onto the sky \citep[e.g.,][]{marion:2014,
tsishchankava:2025} and gets attenuated by spatial filtering
\citep[e.g.,][]{wang:2021} and possibly spectral-bandwidth smearing
\citep[e.g.,][]{zhao:2007a, lachaume:berger:2013}. Furthermore, a detection
requires a sufficient sampling of the spatial Fourier frequencies for
companions with large projected distances \citep{absil:2010}, and sufficient
angular resolution in case of small projected distances.

The system of \object[HD7788A]{\ktuc~A} is uniquely important for the study
of hot exozodis because it has long been thought to host dust whose brightness
varies on a yearly timescale \citep{ertel:2014, ertel:2016}. Constraints on
temporal variability can aid in determining the origin of hot exozodis
\citep[e.g.,][]{pearce:2022b} and hence \ktuc~A is continuously monitored.
The primary component, \ktuc~Aa, is of spectral type F6 with an effective
temperature of $\sim \qty{6500}{K}$ \citep{fuhrmann:2017} and is located at
a distance of $\sim \qty{21}{pc}$ (see Sect.~\ref{subsect_orbit_fit}). Its
suggested age is $\sim \qty{2}{Gyr}$ and it has slightly evolved off the main
sequence \citep{tokovinin:2020}. Together with \object[HD7788B]{\ktuc~B} at a
projected separation of $\sim \qty{5}{as}$, \ktuc~A forms the system of
\ktuc\footnote{From the \texttt{SIMBAD} Astronomical Database \citep{wenger:2000}.}
(HD 7788, HIP 5896) that is gravitationally bound to the binary system HD 7693
(HIP 5842) at \qty{318}{as} separation \citep[for a detailed description
see][]{tokovinin:2020}.

Based on data obtained in the $H$-band with the Precision Integrated-Optics
Near-infrared Imaging Experiment \citep[PIONIER;][]{le_bouquin:2011} at the
Very Large Telescope Interferometer \citep[VLTI;][]{haubois:2020},
\citet{ertel:2014} announced a hot exozodi around \ktuc~Aa based on
observations from 2012 July, after \citet{marion:2014} did not find a companion
within a distance of $\qty{100}{mas}$ using the same data. Subsequently,
\citet{ertel:2016} detected no near-infrared excess that could be interpreted
as a hot exozodi based on observations of 2013 August, but detected an excess
again based on observations from 2014 November. Lastly, based on observations
with the Multi Aperture mid-Infrared Spectroscopic Experiment
\citep[MATISSE;][]{lopez:2022} at the VLTI in 2019 July, a hot exozodi was
detected in the $L$ band, and no obvious companion signal was reported
\citep{kirchschlager:2020}.

However, \citet{tokovinin:2020} announced a previously unknown astrometric
companion to \ktuc~Aa, \ktuc~Ab, based on residuals in the orbital solution
of \ktuc~B, and differences between Hipparcos \citep{van_leeuwen:2007} and
Gaia \citep{gaia_collaboration:2016} proper motions \citep{brandt:2018,
brandt:2019}; hence \ktuc\ is a triple system, that forms together with the
binary HD 7693 a quintuple system. The astrometric companion \ktuc~Ab was
suspected to be of low mass and, assuming a circular orbit, to have a period
of $\sim \qty{20}{yr}$. Despite this analysis using astrometry from Gaia,
\ktuc~A is not listed in the Gaia non-single star catalog
\citep{gaia_collaboration:2023b, holl:2023}.

In this article, we present interferometric detections of the astrometric
companion to \ktuc~Aa, \ktuc~Ab, using MATISSE and GRAVITY
\citep{gravity_collaboration:2017} at the VLTI; we show that the companion is
a low-mass star on a highly eccentric orbit. In Sect.~\ref{sect_observations},
we describe our observations, the data reduction and selection, and discuss
sources of uncertainty for astrometric analysis. After we present the detection
and verify with GRAVITY that MATISSE delivers accurate relative astrometry of
the companion (Sect.~\ref{sect_companion_detection}), we determine the
companion's position over time in our multi-epoch data and tightly constrain
its orbit and spectral type (Sect.~\ref{sect_multiepoch_analysis}). We discuss
the performance in measuring relative astrometry of companions with MATISSE,
the now confirmed triple system of \ktuc, the impact of the new companion on
previous detections of a hot exozodi around \ktuc~Aa, and provide an outlook
into further studies of this peculiar system (Sect.~\ref{sect_discussion}). A
summary concludes this study (Sect.~\ref{sect_summary}), which is solely
dedicated to the analysis of the companion; we refer the investigation of the
hot exozodi to a forthcoming paper. In this article, we use the term \ktuc~A to
refer to the binary system of the components \ktuc~Aa and Ab, and use the terms
\ktuc~Aa and \ktuc~Ab to refer to the individual components if necessary.


\begin{deluxetable*}{l c c c c c c c}
    \tablecaption{VLTI/MATISSE and GRAVITY Observations of \ktuc~A\label{tab_observations}}
    \tabletypesize{\scriptsize}
    \tablecolumns{8}
    \tablehead{
        \colhead{Date} & \colhead{Science OB Execution Time} & \colhead{AT-Configuration} & \colhead{Instrument} &
        \colhead{GRA4MAT} & \colhead{Spectral Resolution} & \colhead{Chopping} & \colhead{Calibrator(s)}
    }
    \startdata
        \phm{aaa}2019 Jul 9     & 08:21:18 -- 08:33:51 & Medium       & MATISSE & No  & LOW-LM & No\phsnm{a}  & HD 3750, HD 8094\\
        \phm{aaa}2019 Jul 11    & 08:51:42 -- 09:06:17 & Medium       & MATISSE & No  & LOW-LM & No\phsnm{a}  & HD 4138, HD 8315\\
        \phm{aaa}2022 Aug 27    & 08:05:35 -- 08:34:27 & Medium       & MATISSE & No & MED-LM & Yes\snm{a}   & HD 2354, HD 1025\\
        \phm{aaa}2022 Oct 28    & 01:18:04 -- 01:56:56 & Medium       & MATISSE & Yes & MED-LM & Yes\snm{a}   & (HD 2354), HD 1025 \\
        \phm{aaa}2022 Oct 29    & 03:14:03 -- 03:41:03 & Medium       & MATISSE & Yes & MED-LM & Yes\snm{a}   & HD 2354, HD 1025 \\
        \phm{aaa}2023 Jun 17    & 10:06:33 -- 10:34:34 & Medium       & MATISSE & Yes & LOW-LM & Yes\phsnm{a} & HD 8810\phsnm{b} \\
        \phm{aaa}2023 Jun 21    & 07:43:15 -- 08:11:59 & Medium-Large & MATISSE & Yes & LOW-LM & Yes\snm{a}   & HD 8810\phsnm{b} \\
        \phm{aaa}2023 Jul 13    & 07:32:38 -- 07:58:49 & Medium       & MATISSE & Yes & LOW-LM & Yes\phsnm{a} & HD 8810\phsnm{b} \\
        \phm{aaa}2023 Aug 13    & 04:50:39 -- 05:18:11 & Medium       & MATISSE & Yes & LOW-LM & Yes\phsnm{a} & HD 8810\phsnm{b} \\
        \phm{aaa}2024 Oct 25    & 06:00:19 -- 06:24:58 & Large        & MATISSE & Yes & LOW-LM & Yes\phsnm{a} & HD 5457, HD 8810 \\
        \phm{aaa}2024 Oct 26    & 02:42:33 -- 03:09:24 & Large        & MATISSE & Yes & LOW-LM & Yes\phsnm{a} & HD 5457, HD 8810 \\
        \phm{aaa}2024 Oct 26    & 06:34:47 -- 07:01:24 & Large        & MATISSE & Yes & LOW-LM & Yes\phsnm{a} & HD 5457\phsnm{b} \\
        \phm{aaa}2024 Oct 29    & 05:41:08 -- 06:13:33 & Small        & MATISSE & Yes & LOW-LM & Yes\phsnm{a} & HD 5457, HD 8810 \\
        \phm{aaa}2024 Oct 29    & 06:39:23 -- 07:04:20 & Small        & MATISSE & Yes & LOW-LM & Yes\phsnm{a} & HD 8810\snm{b} \\
        \phm{aaa}2024 Oct 29/30 & 23:54:39 -- 00:20:17 & Small        & MATISSE & Yes & LOW-LM & Yes\phsnm{a} & HD 5457, HD 8810 \\
        \phm{aaa}2024 Nov 6     & 03:54:09 -- 04:22:19 & Medium       & MATISSE & Yes & LOW-LM & Yes\phsnm{a} & HD 5457, HD 8810 \\
        \phm{aaa}2024 Nov 6     & 04:51:43 -- 05:17:04 & Medium       & MATISSE & Yes & LOW-LM & Yes\phsnm{a} & HD 8810\snm{b} \\
        \phm{aaa}2024 Nov 7     & 00:15:41 -- 00:40:39 & Medium       & MATISSE & Yes & LOW-LM & Yes\phsnm{a} & HD 5457, HD 8810 \\
        \phm{aaa}2024 Nov 23    & 02:56:37 -- 03:27:37 & Large        & MATISSE & Yes & LOW-LM & Yes\phsnm{a} & HD 5457\phsnm{b} \\
        \tableline
        \phm{aaa}2024 Nov 23    & 01:15:23 -- 01:43:51 & Large        & GRAVITY & --  & HIGH   & --           & HD 2354\phsnm{b} \\
        \cutinhead{Discarded observations}
        \phm{aaa}2022 Oct 29    & 00:02:51 -- 00:30:24 & Medium       & MATISSE & Yes & MED-LM & Yes\phsnm{a} & -- \\
        \phm{aaa}2023 Jun 17    & 09:10:34 -- 09:38:22 & Medium       & MATISSE & Yes & LOW-LM & Yes\phsnm{a} & -- \\
        \phm{aaa}2023 Jul 13    & 08:27:10 -- 08:54:03 & Medium       & MATISSE & Yes & LOW-LM & Yes\phsnm{a} & -- \\
        \phm{aaa}2023 Jul 13    & 09:22:24 -- 09:53:26 & Medium       & MATISSE & Yes & LOW-LM & Yes\phsnm{a} & -- \\
    \enddata
    \tablenotetext{a}{Chopped data is not reliable.}
    \tablenotetext{b}{Calibrator observation from the leading observation used.}
    \tablecomments{
        The listed execution times give the time interval in Universal Time (UT) used to execute
        the whole observing block (OB) of the science target. When two calibrators are listed,
        both are used together to determine the instrumental transfer function. Calibrators in
        parentheses are discarded. The VLTI telescope positions for the listed AT-configurations
        are A0-B2-C1-D0 (small), D0-G2-J3-K0 (medium), A0-G2-J2-J3 (medium-large, an intermediate, nonstandard configuration), and A0-G1-J2-K0
        (large).
    }
\end{deluxetable*}

\section{Observations}
\label{sect_observations}

The light collected with the VLTI Auxiliary Telescopes
\citep[ATs,][]{koehler:flebus:2000} and their New Adaptive Optics Module for
Interferometry \citep[NAOMI;][]{woillez:2019} was injected into the
interferometric instruments MATISSE and GRAVITY. Both instruments combine the
light of four telescopes to perform spectrally dispersed measurements of
interferometric fringes for six pairs of telescopes. All observations with
their different set-ups are summarized in Table~\ref{tab_observations},
including observations that failed or are not of sufficient quality for our
analysis; they were performed under ESO programs 0103.C-0725, 109.23HL,
110.24AJ, 111.24S6, 114.27F3, 114.28HE, and 114.28HE.


\subsection{MATISSE}
\label{subsect_obs_matisse}

MATISSE operates in the mid-infrared bands $L$, $M$, and $N$. As the system of
\ktuc~A is too faint to obtain reliable $M$- and $N$-band measurements with
the ATs, we focus completely on the $L$ band. New MATISSE observations were
executed between 2022 August and 2024 November. They are complemented by two
observations obtained in 2019 July by \citet{kirchschlager:2020}. While
observations from 2019 were performed in stand-alone mode (i.e., fringe
coherencing in the $L$ band), the new observations were performed in GRA4MAT
mode \citep{woillez:2024} employing the GRAVITY fringe tracker operating in
the $K$ band \citep{lacour:2019}. Each observation of \ktuc~A was accompanied
by one or two observations of calibrator stars. Those were optimized for
$L$-band calibration and selected from the MDFC catalog
\citep{cruzalebes:2019} with diameters from the JSDC catalog
\citep{chelli:2016}. Each of our MATISSE observations begins with four 1 minute
exposures performed without chopping (hereafter \textit{non-chopped}), each
with a different configuration of the two beam commuting devices
(BCDs).\footnote{The two BCDs switch the light paths of the four telescopes
pairwise and can be positioned either IN the light path or OUT of it,
respectively. Thus, there are four configurations of the BCDs: IN-IN, IN-OUT,
OUT-IN, and OUT-OUT.} Subsequently, eight 1 minute exposures are conducted with
chopping (hereafter \textit{chopped}), which allows for an improved correction
of the thermal background. Four of these are performed with the BCD
configurations IN-IN and OUT-OUT, respectively. Exceptions to this are the
observations of 2019, for which no chopping was performed. Moreover, in 2022,
using GRA4MAT in combination with chopping delivered unreliable data. While
the operation and stability of the GR4MAT mode were improved by early 2023
\citep{nowak:2024, woillez:2024}, a low atmospheric coherence time can still
lead to unreliable chopped data. This was the case for the execution on 2023
June 21. The observations of 2022 were performed in MED-LM spectral resolution
(resolving power $R = \lambda/\delta\lambda$ = 499, with wavelength $\lambda$
and spectral resolution element $\delta\lambda$) while others were performed
with LOW-LM spectral resolution \citep[$R = 31.5$,][Table 2]{lopez:2022} to
maximize the signal-to-noise ratio for a search for hot exozodis.


\subsection{GRAVITY}
\label{subsect_obs_gravity}

The last observation with MATISSE on 2024 November 23 was accompanied by
observations during the same night with GRAVITY, which operates in the
near-infrared $K$ band. These observations were performed in single-field
on-axis mode: the system was bright enough such that the light was split
between the fringe tracker and the spectrograph. For the spectrograph, we
selected HIGH spectral resolution ($R\sim \num{4000}$) to maximize the
coherence length and avoid bandwidth smearing (see
Appendix~\ref{appdx_sect_bandwidth_smearing}). The observation of \ktuc~A was
succeeded by the observation of the calibrator star HD~2354, which we selected
with the \texttt{SearchCal} tool from the Jean-Marie Mariotti Center (JMMC)
that includes the JMDC \citep{chelli:2016, duvert:2016} and JSDC
\citep{chelli:2016} catalogs. Each science and calibrator observation consisted
of \num{24} frames of \qty{30}{s} on target (total of 12 minutes), and \num{12}
frames of \qty{30}{s} on-sky (total of 6 minutes).


\subsection{Data Reduction and Selection}
\label{subsect_data_reduction}

We reduce raw MATISSE data (including archival data from 2019) using the
MATISSE data reduction software \citep[see][]{millour:2016} in version 2.0.2.
To run the reduction and calibrate the reduced data, we use the Python package
\texttt{mat\_tools}.\footnote{\url{https://github.com/Matisse-Consortium/tools}}
Following standard data reduction, the calibrated non-chopped exposures would
be merged into a single 4 minute exposure, the chopped exposures into an 8
minute exposure, both averaging observations with different BCD positions.
However, this causes significant smearing of the companion signal due to sky
rotation. To avoid this, we use the four individual 1 minute non-chopped
exposures. We prefer them over the chopped exposures because their
signal-to-noise ratio is higher, as chopping reduces the time on target, and we
will focus the analysis on the closure phases (see
Sect.~\ref{sect_companion_detection}) that do not profit from chopping and
thermal-background subtraction, as they are derived from the position of the
fringes on the detector. The default merging of a full BCD sequence of
non-chopped exposures corrects the closure phases for instrumental systematic
effects to a precision of $\lesssim \qty{1}{\degree}$. As we keep the
exposures separate, we have to rely on the comparison with calibrator
observations to correct for systematic effects. On 2024 November 23 the BCD
sequence was executed twice, thus resulting in eight individual exposures. On
2019 July 9 the OUT-OUT exposure failed, while on 2019 July 11 the IN-IN
exposure failed, thus resulting in only three exposures for each of the dates.

For the LOW-LM spectral resolution of MATISSE, our analysis encompasses
wavelengths between \qty{3.1}{\um} and \qty{3.9}{\um}, which provides
comprehensive coverage of the $L$ band while avoiding the band edges with
a low signal-to-noise ratio. For the observations from 2019, the lower boundary
is \qty{3.28}{\um} due to a specific instrument setup. For the MED-LM spectral
resolution, we analyze wavelengths between \qty{3.35}{\um} and \qty{3.9}{\um};
the region for wavelengths $< \qty{3.35}{\um}$ is contaminated by several
telluric lines and is particularly noisy for the observation of 2022 August 27.
We refrain from smoothing the MED-LM data down to LOW-LM resolution using a
sliding average as offered by the data reduction software because this process
attenuates the companion signal.

GRAVITY observations were reduced using the GRAVITY data reduction software
\citep{lapeyrere:2014} in version 1.6.6 with standard parameters. As with
MATISSE data, significant smearing of the companion signal occurs if all
frames are combined into one exposure. Therefore, we group four frames into
exposures of 2 minute length and yield six individual data sets. We test
grouping six frames into exposures of 3 minutes, which does not significantly
change the retrieved astrometry of the companion
(Sect.~\ref{subsect_matisse_vs_gravity}) compared to the 2 minute exposures.
This also confirms that 1 minute for MATISSE exposures is short enough to avoid
significant smearing by sky rotation. We analyze wavelengths between
\qty{2.05}{\um} and \qty{2.35}{\um}, avoiding band edges with a low
signal-to-noise ratio. All calibrated Oifits files \citep{duvert:2017} from
both MATISSE and GRAVITY observations that are used for analysis are available
at the Optical Interferometry Database
(OIDB).\footnote{\href{http://oidb.jmmc.fr}{http://oidb.jmmc.fr}}


\subsection{Uncertainty of Spatial Frequencies}
\label{subsect_uncertainty_spat_frequ}

The spatial frequency at which the source brightness distribution is probed
is defined as $B/\lambda$, which is the ratio of the absolute separation
vector of two telescopes projected onto the sky (the baseline $B$) and the
observing wavelength $\lambda$. Therefore, the finite precision on the
baselines and wavelengths leads to uncertainties of the spatial frequencies
that limit the precision of the astrometric analysis. However, it is neither
considered by the MATISSE nor the GRAVITY data reduction software.

The relative uncertainty of the wavelength calibration for GRAVITY in HIGH
spectral resolution is $\Delta \lambda / \lambda = \num{2e-4}$
\citep{gallenne:2023}, based on daily calibration using the internal laser
metrology \citep{gillessen:2012}. For MATISSE in MED-LM and HIGH-LM
resolution, the wavelength calibration is precise to the extent of one
detector pixel (MATISSE Consortium 2025, private
communication).\footnote{This information is contained in the commissioning
document VLT-TRE-MAT-15860-9141 v1.0.}
One resolution element of the MATISSE LM-band arm is sampled by approximately
five pixels. Thus, assuming a wavelength of \qty{3.5}{\um} centered in the
$L$ band, in case of the MED-LM resolution, 1 pixel covers a wavelength
interval of $\approx \qty{0.0014}{\um}$ which results in
$\Delta \lambda / \lambda \approx \num{4e-4}$.

To verify this behavior for the LOW-LM resolution, we analyze \num{38}
wavelength calibrations between 2019 July 26 and 2024 September 10 and find
that the standard deviation of the wavelength associated with a single pixel
is $\approx \qty{0.019}{\um}$. This is slightly smaller than the wavelength
coverage of one pixel in the central $L$ band of $\approx \qty{0.022}{\um}$.
Therefore, we confirm that the wavelength calibration is precise to the extend
of one detector pixel also in LOW-LM resolution and adopt this for further
analysis. This results in a relative wavelength uncertainty for LOW-LM of
$\Delta \lambda / \lambda \approx \num{6.3e-3}$.

To estimate the baseline precision, we consider the \textit{imaging baselines}
\citep{woillez:lacour:2013}. The estimated relative precision of the VLTI
imaging baselines is \num{5e-5} (ESO 2025, private communication), which is
already smaller than the wavelength uncertainty of GRAVITY. Additionally, this
uncertainty applies to each baseline individually. That is, some baseline
lengths are underestimated, while others are overestimated, which further
mitigates the impact in a joint analysis of data from all telescope pairs.
Therefore, we neglect the baseline uncertainty, and thus the relative
uncertainty of the spatial frequencies is given by the relative wavelength
uncertainty $\Delta \lambda / \lambda$.


\section{Detection of a High-contrast Companion}
\label{sect_companion_detection}

The closure phases from both MATISSE and GRAVITY show an oscillating signal
(see Figs.~\ref{fig_cp_matisse}, \ref{fig_cp_gravity}), indicating that the
source brightness distribution is not point-symmetric. A stellar companion to
the primary star would cause such an oscillating signal, with the amplitude
being related to the binary flux ratio \citep[e.g.,][]{le_bouquin:absil:2012,
tsishchankava:2025}. Given that an astrometric companion to \ktuc~Aa has
already been inferred \citep{tokovinin:2020}, we assume a stellar companion,
\ktuc~Ab, to be the source of the asymmetry. This assumption is validated in
retrospect by the analysis of the whole observational sample
(Sect.~\ref{sect_multiepoch_analysis}).

A companion would also cause an oscillating signal in the (squared)
visibilities, which is present in our data. However, visibilities are affected
by a uniform circumstellar dust distribution that causes a visibility deficit
compared to the signal from the stellar photosphere alone
\citep[e.g.,][]{absil:2006, di_folco:2007}. Such a dust distribution is
point-symmetric and thus has no effect on the closure phases. As the companion
signal is clearly pronounced in the closure phases, visibilities are not needed
to improve a marginal companion detection \citep[e.g.,][]{marion:2014}, and we
restrict our analysis to the closure phases to reduce bias by possible
circumstellar dust.

Few investigations were dedicated to detecting high-contrast companions with
MATISSE. \citet{lopez:2022} demonstrated this capability during commissioning,
observing a known binary with a flux ratio $f$ from companion to primary of
$\sim \qty{2}{\percent}$, while \citet{varga:2025} detected a new companion,
albeit with a larger flux ratio of $f \sim \qty{10}{\percent}$. However, the
detections of high-contrast companions and their relative astrometry retrieved
from MATISSE observations have not been verified independently. For this
purpose, we use the joint GRAVITY and MATISSE observations from 2024 November
23.


\subsection{Companion Search and Detection}
\label{subsect_companion_search}

\begin{figure}
    \centering
    \includegraphics[width=1\linewidth]{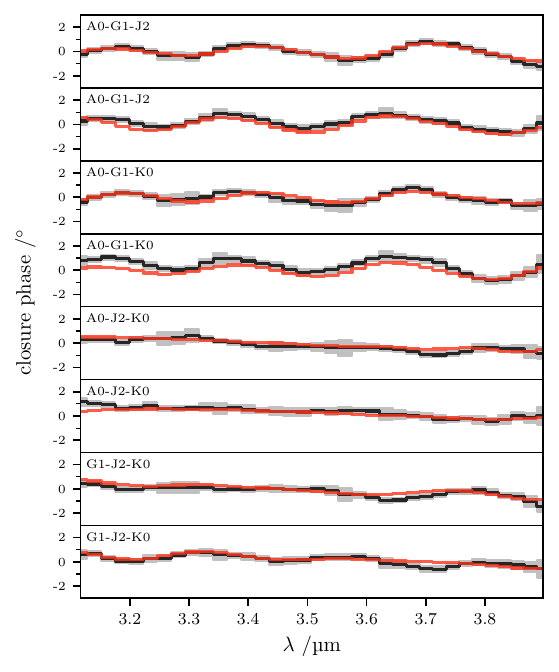}
    \caption{
        MATISSE closure phases obtained on 2024 November 23 in BCD position
        OUT-OUT (black) with uncertainties in gray. The BCD cycle was executed
        twice, delivering two measurements for each telescope triplet, which
        is denoted in each panel by the combination of AT positions (e.g.,
        A0-G1-J2 for the positions A0, G1, and J2). The two BCD cycles were
        recorded with a pause of 8 minutes in between, during which the
        chopping sequence was performed. Due to sky rotation during this
        period, the measurements deviate slightly from each other. A binary
        model was fitted to the data of all four BCD positions (of which
        three are not shown for clarity of illustration) with the best-fit
        model shown in red.
    }
    \label{fig_cp_matisse}
\end{figure}
\begin{figure}
    \centering
    \includegraphics[width=1\linewidth]{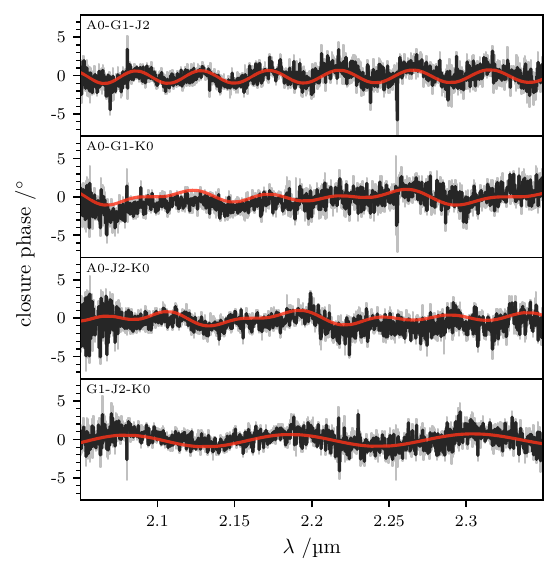}
    \caption{
        GRAVITY closure phases from the observation of 2024 November 23 for
        the first of six 2 minute exposures (black) with uncertainties in
        gray. Each panel belongs to one telescope triplet. A binary model
        was fitted to the data of all six exposures (of which five are not
        shown for clarity of illustration) with the best-fit model shown in
        red.
    }
    \label{fig_cp_gravity}
\end{figure}

We search for the signature of a companion to \ktuc~Aa separately in the
MATISSE and GRAVITY closure phases obtained on 2024 November 23 using the
numerical modeling suite \texttt{PMOIRED} v1.3.1 \citep{merand:2022,
merand:2024}. We model the system with a central uniform disk with a diameter
of \qty{0.739}{mas} \citep{ertel:2014} for the primary and an off-axis point
source for the companion. For a companion search in interferometric data, the
$\chi^2$-map for the companion position typically shows several local minima
\citep[e.g.,][]{absil:2011, marion:2014, gallenne:2015}. Thus, we use the
\texttt{CANDID} grid search algorithm \citep[embedded in
\texttt{PMOIRED},][]{gallenne:2015} with a grid spacing of \qty{2}{mas} to find
the global best-fit solution. \texttt{PMOIRED} takes the effect of
spectral-bandwidth smearing into account, which affects our MATISSE
observations with LOW-LM spectral resolution (see
Appendix~\ref{appdx_sect_bandwidth_smearing}, including how bandwidth smearing is
treated by \texttt{PMOIRED}).

Measurements from the VLTI are highly correlated \citep[e.g.,][]{lachaume:2019,
lachaume:2021}. \citet{ertel:2014} determined that for squared visibilities,
the dominating source of correlation is among the spectrally dispersed data
belonging to a certain baseline. We extend this finding to the closure phases
and assume all closure phases belonging to a certain telescope triplet are
fully correlated. We compute the final fit results using a bootstrapping
procedure \citep{efron:1979, efron:1982} with the initial fit parameters set to
the global solution from the grid search. The bootstrapping routine accounts
for the correlation by keeping all data belonging to a certain triplet for a
certain exposure together, thus sampling triplets instead of individual data
points. We draw \num{2000} data samples in each bootstrapping fit, which is
sufficient to yield stable results. We take the final best-fit parameters to be
the median values from the \num{2000} results and compute their uncertainties
by averaging the differences between the median and the 0.16 and 0.84
quantiles, respectively. This procedure yields similar best-fit values
compared to the global solution from the \texttt{CANDID} grid search, but
increases uncertainties.

We detect the signal of a companion in the MATISSE closure phases (see
Fig.~\ref{fig_cp_matisse}). Assessing the numerical significance of a
detection is challenging due to the correlation among the data. However,
considering that the binary flux ratio obtained from the closure phases
$f\sbs{CP} = \qty{1.01(0.06)}{\percent}$ is highly significant and the model
reproduces well the closure phase oscillations across telescope triplets, we
consider this a highly significant detection. Analyzing the GRAVITY closure
phases, we detect the companion signal as well and confirm the MATISSE
detection. The flux ratio of $f\sbs{CP} = \qty{0.70(0.02)}{\percent}$ is again
highly significant, and the oscillations are well reproduced (see
Fig.~\ref{fig_cp_gravity}).

\begin{deluxetable*}{l c c c c c c c c}
    \tablecaption{Position and Flux of the Companion \ktuc~Ab Relative to the Primary Star \ktuc~Aa\label{tab_astrometry}}
    \tabletypesize{\scriptsize}
    \tablecolumns{7}
    \tablehead{
        \colhead{Date} & \colhead{MJD} & \colhead{$\rho$ /\unit{mas}} & \colhead{$\epsilon_\rho$ /\unit{mas}} &
        \colhead{PA /\unit{\degree}} & \colhead{$\epsilon\sbs{PA}$ /\unit{\degree}} & 
        \colhead{$f\sbs{CP}$ /\unit{\percent}} & \colhead{$f$ /\unit{\percent}} & \colhead{Instrument}
    }
    \startdata
        \phm{a}2019 Jul 9 and 11  & 58674.366 & \num{326.2(2.2)}    & \phm{$-$}\num{8.1}\phn & \num{233.62(0.10)}  & \num{-0.78}         &  \num{0.51(0.05)} & \num{1.29(0.13)} & MATISSE \\
        \phm{a}2022 Aug 27    & 59818.347     & \num{228.1(0.2)}    & \num{-0.2}\phn         &  \num{217.10(0.04)} & \phm{$-$}\num{0.04} &  \num{0.95(0.04)} & \num{1.47(0.07)} & MATISSE \\
        \phm{a}2022 Oct 28    & 59880.071     & \num{211.0(0.2)}    & \phm{$-$}\num{0.03}    &  \num{215.29(0.07)} & \num{-0.07}         &  \num{0.90(0.06)} & \num{1.3(0.09)}  & MATISSE \\
        \phm{a}2022 Oct 29    & 59881.143     & \num{210.7(0.2)}    & \phm{$-$}\num{0.03}    &  \num{215.27(0.05)} & \num{-0.06}         &  \num{0.81(0.03)} & \num{1.18(0.05)} & MATISSE \\
        \phm{a}2023 Jun 17    & 60112.433     & \num{112.1(0.7)}    & \num{-0.4}\phn         &  \num{203.34(0.17)} & \phm{$-$}\num{0.04} &  \num{0.87(0.10)} & \num{0.97(0.11)} & MATISSE \\
        \phm{a}2023 Jun 21    & 60116.332     & \num{109.7(0.7)}    & \num{-0.3}\phn         &  \num{202.75(0.12)} & \num{-0.13}         &  \num{1.28(0.06)} & \num{1.41(0.06)} & MATISSE \\
        \phm{a}2023 Jul 13    & 60138.324     & \phn\num{94.3(0.6)} & \num{-0.7}\phn         &  \num{200.78(0.16)} & \phm{$-$}\num{0.65} &  \num{1.52(0.15)} & \num{1.63(0.16)} & MATISSE \\
        \phm{a}2023 Aug 13    & 60169.213     & \phn\num{69.7(0.5)} & \phm{$-$}\num{0.1}\phn &  \num{193.37(0.27)} & \num{-0.46}         &  \num{1.26(0.07)} & \num{1.31(0.07)} & MATISSE \\
        \phm{a}2024 Oct 25    & 60608.260     & \num{164.2(1.0)}    & \num{-0.7}\phn         &  \num{257.00(0.03)} & \num{-0.01}         &  \num{1.06(0.03)} & \num{1.32(0.04)} & MATISSE \\
        \phm{a}2024 Oct 26    & 60609.124     & \num{164.2(1.0)}    & \num{-1.0}\phn         &  \num{257.09(0.03)} & \phm{$-$}\num{0.09} &  \num{1.26(0.05)} & \num{1.57(0.06)} & MATISSE \\
        \phm{a}2024 Oct 26    & 60609.285     & \num{163.8(1.0)}    & \num{-1.2}\phn         &  \num{256.96(0.02)} & \phm{$-$}\num{0.01} &  \num{1.17(0.03)} & \num{1.45(0.03)} & MATISSE \\
        \phm{a}2024 Oct 29    & 60612.251     & \num{165.4(1.1)}    & \num{-0.7}\phn         &  \num{256.59(0.07)} & \num{-0.20}         &  \num{0.99(0.04)} & \num{1.24(0.04)} & MATISSE \\
        \phm{a}2024 Oct 29    & 60612.287     & \num{164.8(1.1)}    & \num{-1.0}\phn         &  \num{256.62(0.09)} & \num{-0.29}         &  \num{1.01(0.05)} & \num{1.27(0.07)} & MATISSE \\
        \phm{a}2024 Oct 29/30 & 60613.006     & \num{165.2(1.1)}    & \num{-0.9}\phn         &  \num{256.81(0.10)} & \phm{$-$}\num{0.04} &  \num{1.11(0.02)} & \num{1.39(0.02)} & MATISSE \\
        \phm{a}2024 Nov 6     & 60620.174     & \num{167.0(1.1)}    & \num{-1.2}\phn         &  \num{256.47(0.06)} & \phm{$-$}\num{0.03} &  \num{1.2(0.06)}  & \num{1.52(0.08)} & MATISSE \\
        \phm{a}2024 Nov 6     & 60620.212     & \num{167.1(1.1)}    & \num{-1.3}\phn         &  \num{256.16(0.11)} & \num{-0.20}         &  \num{0.69(0.09)} & \num{0.87(0.11)} & MATISSE \\
        \phm{a}2024 Nov 7     & 60621.021     & \num{167.2(1.1)}    & \num{-1.3}\phn         &  \num{256.45(0.04)} & \phm{$-$}\num{0.01} &  \num{1.04(0.04)} & \num{1.31(0.05)} & MATISSE \\
        \phm{a}2024 Nov 23    & 60637.132     & \num{171.5(1.1)}    & \num{-1.1}\phn         &  \num{255.71(0.09)} & \phm{$-$}\num{0.04} &  \num{1.01(0.06)} & \num{1.29(0.08)} & MATISSE \\
        \tableline
        \phm{a}2024 Nov 23    & 60637.063     & \num{172.66(0.04)}  & \num{0.04}             &  \num{255.64(0.02)} & \num{-0.04} &  \num{0.70(0.02)} & \num{1.87(0.05)} & GRAVITY \\
    \enddata
    \tablecomments{
        The MJD of a certain epoch is computed by averaging the MJDs of all corresponding subexposures. The 2019
        epoch was excluded from the orbit analysis.
    }
\end{deluxetable*}


\vspace{-1cm}
\subsection{Astrometry: MATISSE versus GRAVITY}
\label{subsect_matisse_vs_gravity}

Besides the binary flux ratio, the model fit predicts the companion position
relative to the primary, expressed by the position angle (PA),\footnote{The
primary star is in the coordinate center, $\textrm{PA} = \qty{0}{\degree}$ is
to the north and increases counterclockwise.} and the separation $\rho$. We
find for the position angle $\textrm{PA} = \qty{255.71(0.09)}{\degree}$ from
MATISSE and $\textrm{PA} = \qty{255.64(0.02)}{\degree}$ from GRAVITY, showing
excellent agreement. In the case of the separation $\rho$, the relative
uncertainty of the spatial frequencies, given here by the relative uncertainty
of the wavelength $\Delta\lambda/\lambda$
(Sect.~\ref{subsect_uncertainty_spat_frequ}), acts as a scaling factor and limits
its precision. We compute the final uncertainty of $\rho$ by quadratically
adding the contributions from the bootstrapping fit $\Delta\rho\sbs{fit}$ and
the relative uncertainty of the wavelengths, thus
\begin{equation}
    \Delta\rho = \sqrt{\left( \Delta\rho\sbs{fit} \right)^2
                 + \left( \frac{\Delta\lambda}{\lambda}\rho \right)^2} \; .
\end{equation}
We find from MATISSE $\rho = \qty{171.5(1.1)}{mas}$ and from GRAVITY
$\rho = \qty{172.66(0.04)}{mas}$. The difference between the two results for
$\rho$ is only marginally larger than the uncertainty of the less precise value
from MATISSE. We conclude that GRAVITY and MATISSE deliver consistent
measurements, and the astrometry retrieved from MATISSE observations is robust
to milliarcsecond precision.


\section{Multi-epoch Analysis}
\label{sect_multiepoch_analysis}

Using the same procedure as discussed in Sect.~\ref{subsect_companion_search},
we detect a companion accompanying the primary star in all MATISSE observations
and determine its flux ratio and position relative to the primary star
(Table~\ref{tab_astrometry}).


\subsection{Orbit Fit}
\label{subsect_orbit_fit}

The companion positions are reasonable for all epochs except the two
observations from 2019: when analyzed individually, the results are
inconsistent with each other. When analyzed jointly with a grid search region
primed to the predictions based on all other epochs, we find a position
consistent with the predictions that we list in Table~\ref{tab_astrometry}.
However, this position is sensitive to small changes in the analyzed wavelength
interval and changes up to $\sim \qty{20}{mas}$. The analysis of the 2019 data
generally suffers from a smaller wavelength coverage, leading to fewer spatial
frequencies being probed and only three instead of four successful 1 minute
exposures (see Sect.~\ref{subsect_data_reduction}). Furthermore, the companion
signal for this large separation of $\gtrsim \qty{300}{mas}$ is strongly
attenuated by both bandwidth smearing and spatial filtering (see later in this
section and Appendix~\ref{appdx_sect_matisse_attenuation}). Therefore, we exclude
the 2019 epoch in the subsequent orbit analysis.

To constrain the physical extension of the orbit, the distance to the system is
required. The astrometry solution of \ktuc~A from Gaia DR3
\citep{gaia_collaboration:2016, gaia_collaboration:2023a} is unreliable. This
is indicated by its RUWE value of $\approx \num{10.4}$, which should be
$\simeq \num{1}$ for a reliable solution \citep{lindegren:2021}. Thus, we
apply the reliable solution of \ktuc~B with $\textrm{RUWE} \approx \num{1.2}$.
Its parallax of \qty{47.65(0.02)}{mas} translates to a distance of
\qty{20.986(0.009)}{pc}.

To constrain the companion's orbital parameters, we use the numerical tool
\texttt{orbitize!} v3.1.0 \citep{blunt:2020, blunt:2024}. The parameter space
is explored via the Markov chain Monte Carlo (MCMC) method with parallel
tempering \citep{swendsen:wang:1986, geyer:1991, earl:deem:2005} using the
embedded tool \texttt{ptemcee} \citep{vousden:2016, vousden:2021} that is
based on \texttt{emcee} \citep{foreman-mackey:2013}; see
Appendix~\ref{appdx_sect_MCMC} for details on the MCMC setup such as the
priors and the corner plot.

\begin{figure*}
    \centering
    \includegraphics[width=1\linewidth]{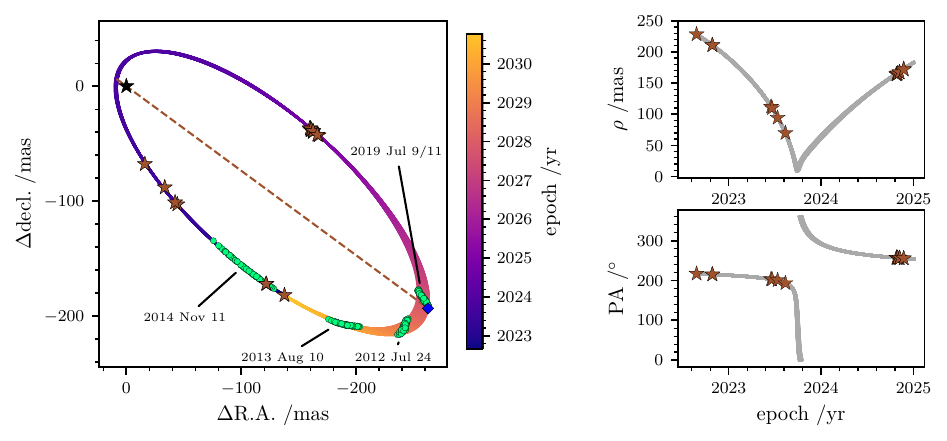}
    \caption{
        One hundred orbits randomly selected from the posterior distribution
        computed by an MCMC fit to relative astrometry (brown stars).
        Left: orbits show the displacement of the companion \ktuc~Ab in decl.
        and R.A. ($\Delta$decl., $\Delta$R.A.) with the primary star \ktuc~Aa
        in the coordinate center (black star). Colors indicate the position of
        \ktuc~Ab along the orbits, starting with our first detection on 2022
        August 27, and ending after one orbital period. The dashed brown line
        connects the peri- and apoastron for the best-fit orbital parameters.
        The blue diamond indicates the tentative position for 2019 July 9/11,
        which was not used to constrain the orbit. In several past PIONIER and
        MATISSE observations, \ktuc~Ab was not detected (see
        Sect.~\ref{subsect_nondetections}) and green points show for the
        selected orbits its predicted past positions at these epochs that are
        denoted by text insets.
        Right: gray curves show for the same orbits as on the left the
        companion separation $\rho$ and the PA of \ktuc~Ab with
        $\textrm{PA} = \qty{0}{\degree}$ to the north and increasing
        counterclockwise. Figure created with a modified \texttt{orbitize!}\
        plot routine.
    }
    \label{fig_orbit}
\end{figure*}

We achieve fully converged chains and show a selection of \num{100} orbits from
the posterior distribution in Fig.~\ref{fig_orbit}. We take the best-fit
orbital parameters to be the median values from the posterior distribution. We
take the uncertainties as the differences between the median and the 0.16 and
0.84 quantiles, respectively, which hence enclose \qty{68}{\percent} of the
distributed values (Table~\ref{tab_orbit}). There are measurements of the
radial velocity of \ktuc~Aa \citep{nordstroem:2004, tokovinin:2015a,
tokovinin:2015b, tokovinin:2020, fuhrmann:2017}, but no radial velocity signal
due to the companion was reported by \citet{tokovinin:2020}. Without
constraining radial velocity data of \ktuc~Aa, we cannot determine the movement
of the companion along the line of sight and hence the results for the argument
of periastron $\omega$ and the position angle of ascending node $\Omega$ are
degenerate to a shift of \qty{180}{\degree} \citep[e.g.,][]{beust:2014}. The
dimensionless quantity $\tau$ parametrizes the periastron passage and is
defined by $\tau = (t\sbs{p} - t\sbs{ref})/P\sbs{orb}$, where $t\sbs{ref}$ is
an arbitrary reference date, \qty{58849}{MJD} (2020 January 1) for
\texttt{orbitize!}\ default settings, $t\sbs{p}$ is the next time of periastron
passage after $t\sbs{ref}$, and $P\sbs{orb}$ is the orbital period computed by
Kepler's third law. Further parameters are the combined mass of primary and
companion $M\sbs{tot}$, and the orbit's semi-major axis $a$, eccentricity $e$,
and inclination $i$.

The well-constrained orbit of the companion verifies that the off-axis point
source detected in various epochs belongs to the same astronomical object that
is not a moving background source but is gravitationally bound to \ktuc~Aa.
The orbit is highly eccentric with $e = \num{0.9356}$. While the apoastron is
at a distance from the primary of $\approx \qty{9.3}{au}$, the periastron is at
only $\approx \qty{0.3}{au}$, approximately at $\sim \num{40}$ stellar radii
of \ktuc~Aa (based on the angular diameter and parallax). This coincides with
the location of hot dust as deduced by \citet{kirchschlager:2020}. The most
recent periastron passage occurred approximately between 2023 October 1 and 5
(see Table~\ref{tab_orbit} for $t\sbs{p}$). The next periastron passage will
occur approximately between 2031 October 14 and 2032 January 2; precise
measurements of the companion's location near the passage will further
constrain the orbit \citep[e.g.,][]{idel:2025}. In addition, during the
passage, the radial velocity signal of \ktuc~Aa will be well detectable with
various instruments; the semi-amplitude is $K \sim \qty{8}{\km\per\s}$.
Determining the sign of the radial velocity during the passage will break the
degeneracy of $\omega$ and $\Omega$.

We compute the residuals of the orbit solution for the companion separation
$\epsilon_\rho$ and the PA $\epsilon\sbs{PA}$ by respectively subtracting the
values predicted by the best-fit orbital model from those deduced from the
closure phases (Table~\ref{tab_astrometry}). Excluding the epoch of 2019, which
was not used for the modeling, the residuals of the separation $\epsilon_\rho$
are on the order of the estimated uncertainties $\Delta\rho$. The residuals of
the position angle $\epsilon\sbs{PA}$ are up to several times larger than the
estimated uncertainties (see Sect.~\ref{subsect_astrometry_with_matisse}).

\begin{deluxetable}{cc}
    \tablecaption{Parameters of the Orbit of the Companion \ktuc~Ab
    from the MCMC fit\label{tab_orbit}}
    \tablewidth{0pt}
    \tabletypesize{\normalsize}
    \tablehead{\colhead{Parameter} & \colhead{Value} }
    \tablecolumns{2}
    \startdata
        $a$ /\unit{au}           & $4.822 \substack{+0.077 \\ -0.071}$ \\[3pt]
        $e$                      & $0.9356 \substack{+0.0024 \\ -0.0024}$ \\[3pt]
        $i$ /\unit{\degree}      & $126.13 \substack{+0.80 \\ -0.83}$  \\[3pt]
        $\omega$ /\unit{\degree} & $(120.9 \ \mathrm{or} +180) \substack{+1.5 \\ -1.4}$ \\[3pt]
        $\Omega$ /\unit{\degree} & $(189.0 \ \mathrm{or} +180) \substack{+1.1 \\ -1.0}$ \\[3pt]
        $\tau$                   & $0.4611 \substack{+0.0057 \\ -0.0057}$ \\[3pt]
        $t\sbs{p}$ /MJD          & $\num{60219.8} \substack{+1.8 \\ -1.8}$ \\[3pt]
        $P\sbs{orb}$ /\unit{yr}  & $8.14 \substack{+0.11 \\ -0.10}$ \\[3pt]
        $M\sbs{tot}$ /$M_\odot$      & $1.69 \substack{+0.11 \\ -0.10}$
    \enddata
    \tablecomments{
        The parameter $M\sbs{tot}$ represents the combined mass of \ktuc~Aa and Ab.
    }
\end{deluxetable}

\subsection{Companion Brightness and Spectral Type}
\label{subsect_brightness_spectral_type}

For both MATISSE and GRAVITY, the flux of an off-axis source, such as the
detected companion, gets attenuated compared to a centered source by spatial
filtering. We correct the thereby biased flux ratio of companion to primary
star $f\sbs{CP}$ as derived from the closure phases to yield the nonattenuated
flux ratio $f$ (Table~\ref{tab_astrometry}). In the case of MATISSE, which
employs pinholes for spatial filtering, we describe and verify the resulting
attenuation for the first time in Appendix~\ref{appdx_sect_matisse_attenuation}.
The final $L$-band flux ratio is obtained by computing the nonweighted average
of all corrected values of $f$ to $f_L = \qty{1.32(0.04)}{\percent}$ with the
uncertainty being the standard error of the mean.

The dependence of the flux ratio $f\sbs{CP}$ on the companion separation $\rho$
in the $L$ band can be explained solely by the instrument characteristics.
Thus, the companion's intrinsic brightness is constant with separation, and
hence it has to be self-luminous; an externally heated object such as a dust
clump would change temperature and thus color and luminosity along its
eccentric orbit because the cooling timescale is much shorter than the orbital
timescale \citep{pearce:2022b, bensberg:wolf:2022}. Together with the
well-constrained orbit, this confirms that the off-axis source is a stellar
companion to \ktuc~Aa as we assumed: \ktuc~Ab.

To estimate the spectral type of this self-luminous companion \ktuc~Ab, we use
the dynamical mass of the system. The mass estimate for \ktuc~Aa is
\qty{1.36}{\Msun}, derived by \citet{fuhrmann:2017} using the $V$-band
magnitude of \ktuc~A; by using it, we neglect any contamination by \ktuc~Ab in
the $V$-band measurement. Subtracting the mass of \ktuc~Aa from the total mass
$M\sbs{tot}$ yields $\sim \qty{0.33(0.10)}{\Msun}$ for the mass of \ktuc~Ab.
This indicates a spectral type between M2.0$\,$V and M5.0$\,$V
\citep[][Table~6]{cifuentes:2020}.

In contrast to MATISSE, GRAVITY uses single-mode optical fibers for spatial
filtering. When using eq.\ A4 of \citet[][Appendix~A]{wang:2021} to correct for
spatial filtering, assuming a wavelength of \qty{2.2}{\um} and a telescope
diameter of \qty{1.8}{\meter} for the ATs, we yield a $K$-band flux ratio of
$f_K = \qty{1.87(0.05)}{\percent}$ from our single observation (see
Sect.~\ref{subsect_companion_search}). This is larger than the flux ratio in
the $L$-band, which contradicts the fact that \ktuc~Aa is of spectral type F
and \ktuc~Ab is of type M. However, the equation of \citet{wang:2021} assumes
the source being perfectly centered on the fiber and no wave front disturbances
due to atmospheric effects. The flux ratios determined from our MATISSE
observations scatter considerably around the theoretically expected values
(Fig.~\ref{fig_spatial_attenuation}), likely due to imperfect centering and
atmospheric effects. Thus, we conclude that with only one GRAVITY observation,
we cannot reliably determine the $K$-band flux ratio. However, GRAVITY and
MATISSE use individual mechanisms to center the source, and an assessment of
GRAVITY’s behavior for systems with companion separations
$\rho \gtrsim \qty{200}{mas}$ and different binary brightness ratios, both
under different atmospheric conditions, is needed.

Instead of the flux ratio in the $K$ band, we use the measured ratio in the $L$
band of \qty{1.32}{\percent} and PHOENIX synthetic spectra \citep[HiRes spectra
with solar metallicity,][]{husser:2013}\footnote{
\href{https://phoenix.astro.physik.uni-goettingen.de/}{
https://phoenix.astro.physik.uni-goettingen.de/}} to estimate the $K$-band flux
ratio. For \ktuc~Aa, we utilize a spectrum for an effective temperature of
\qty{6500}{\K} and a logarithmic surface gravity of $\log(g) = \num{4.0}$. For
\ktuc~Ab, we explore spectra for intermediate-type M dwarfs with effective
temperatures between \qty{3000}{\K} and \qty{3400}{\K} and
$\log(g) = \num{5.0}$. Given the measured $L$-band flux ratio, the $K$-band
flux ratio would be $\approx \qty{1.1}{\percent}$ consistent across the
explored spectra. Given the companion separation and the measured flux ratio
determined from the closure phases (Table~\ref{tab_astrometry}), this is
consistent with a fiber offset of \qty{45}{mas} from \ktuc~Aa in the direction
of \ktuc~Ab. An imperfectly corrected wave front would further reduce this
required offset because the fiber injection of the on-axis primary would
decrease while the injection of the off-axis companion would increase.

We use this $K$-band flux ratio of \qty{1.1}{\percent} to do a flux-based
determination of the spectral type, neglecting differences between the
photometric bands $K$ and $K$\sbs{s}. The apparent magnitude of the multiple
system \ktuc\ in the $K$\sbs{s} band from the Two Micron All Sky Survey
\citep[2MASS][]{skrutskie:2006} is $m_\textrm{$K$\sbs{s},$\,$AB} = 3.876$
\citep{skrutskie:2003, cutri:2003}; this is likely the joint magnitude of
\ktuc~Aa, Ab, and B as the latter does not have its own 2MASS identifier and
the component \ktuc~Ab was only recently inferred. Thus, we first correct for
the magnitude of \ktuc~B and find for the binary \ktuc~A the magnitude
$m_\textrm{$K$\sbs{s},$\,$A} \approx \num{4.15}$ (see
Appendix~\ref{appdx_sect_magnitude}). Using the flux ratio of \ktuc~Ab to Aa,
$f_K \approx \qty{1.1}{\percent}$, this results in an apparent magnitude of
\ktuc~Ab of $m_\textrm{$K$\sbs{s},$\,$Ab} \approx \num{9.1}$ and an absolute
magnitude of $M_\textrm{$K$\sbs{s},$\,$Ab} \approx \num{7.4}$. This indicates
a spectral type between M3.5$\,$V and M4.5$\,$V
\citep[][Table~7]{cifuentes:2020}. The corresponding mass from their Table~6
agrees with the mass of $\approx \qty{0.24}{\Msun}$ estimated with the
mass-luminosity relation of \citet[][eq.\ 11]{benedict:2016} and the estimate
from our dynamical mass. Therefore, the mass and flux-based estimates of the
spectral type agree, supporting the consistency of our analysis.

At most times, \ktuc~Ab is located at separations of
$\rho \gtrsim \qty{200}{mas}$ to \ktuc~Aa. With a contrast to the primary of
$\sim 10^{-2}$, it is easily detectable by high-contrast facilities such as the
Very Large Telescope instruments SPHERE \citep{beuzit:2019}, HiRISE
\citep{vigan:2024}, or ERIS \citep{davies:2023, hayoz:2025} that can measure
more precise fluxes across different photometric bands. This would allow for a
more precise determination of the spectral type.


\section{Discussion}
\label{sect_discussion}


\subsection{Astrometric Measurements with MATISSE}
\label{subsect_astrometry_with_matisse}

We interferometrically detected for the first time a stellar companion to
\ktuc~Aa, \ktuc~Ab, that was previously inferred by astrometry
\citep{tokovinin:2020}. Using the combined mass of the stars and their flux
ratio, we constrained the spectral type of \ktuc~Ab to be between M3.5$\,$V
and M4.5$\,$V. The companion signature in the closure phases is mostly smaller
than $\pm \qty{2}{\degree}$ and the smallest inferred binary flux ratio in the
$L$ band based on this signal is $\sim \qty{0.5}{\percent}$. The true flux
ratio derived from the whole observational sample after correcting for spatial
filtering is $\qty{1.32(0.04)}{\percent}$. This makes \ktuc~Ab, to our
knowledge, the stellar companion with the highest contrast ever detected using
MATISSE closure phases. The companion is also detected in most of our (squared)
visibility measurements, but we refrained from including them in the analysis
to prevent any possible circumstellar material, such as a hot-dust
distribution, from degrading the retrieved relative astrometry.

Using GRAVITY, we validated the relative astrometry retrieved from MATISSE
closure phases and verified that MATISSE is able to deliver milliarcsecond
precision on a companion's location in LOW-LM spectral resolution
($R = \num{31.5}$). With this spectral setting, the main source of uncertainty
is the limited precision of the wavelength calibration that directly affects
the retrievable precision of the companion separation. Furthermore, depending
on the employed baselines and the separation, bandwidth smearing can attenuate
the companion's visibility and closure phase signal. Employing higher spectral
resolutions than LOW-LM lifts these issues. However, this decreases the
signal-to-noise ratio, which can be detrimental for observations of faint
primary stars and high-contrast companions.

Based on our best-fit orbital solution, the residuals of the PA are up to
several times larger than the estimated uncertainties (see
Sect.~\ref{subsect_orbit_fit}). This suggests that the uncertainties of the
PAs are underestimated. Possibly, there are systematic effects such as the
discussed efficiency of centering the source within the spatial filter (see
Sect.~\ref{subsect_brightness_spectral_type}), or single observations are
insufficient to estimate the uncertainties via bootstrapping due to the high
correlation among the data (see Sect.~\ref{subsect_companion_search}). For the
companion separation, both effects could be masked as the uncertainty of the
spatial frequencies dominates the error budget for most of the epochs.
\clearpage


\subsection{The Triple System of \ktuc}
\label{subsect_triple_system}

\citet{tokovinin:2020} fitted the orbit of \ktuc~B around \ktuc~A and
identified residuals in agreement with the proper motion anomaly of \ktuc~A
from \citet{brandt:2018}, which they interpreted as signs of an astrometric
companion to \ktuc~Aa. Based on the assumption of a circular orbit of this
companion, \citet{tokovinin:2020} presented a companion mass of
\qty{0.2}{\Msun} as a plausible estimate, but noted that the PA residuals for
\ktuc~B did not suggest a circular orbit. We find for the interferometrically
detected companion a highly eccentric orbit and a mass of
\qty{0.33(10)}{\Msun}, close to the estimate of \citet{tokovinin:2020}, while
using the same mass estimate for \ktuc~Aa. Thus, we conclude that the detected
companion and the inferred astrometric companion are the same object, \ktuc~Ab.

However, our orbit constraints predict that the binary \ktuc~Aa, Ab experienced
periastron passage approximately at the time of the Gaia observation of
DR2.\footnote{Characteristic epoch 2015.62062 for decl., 2015.65609 for R.A.,
see \citet{brandt:2018, brandt:2019}.} Thus, the measured positions of \ktuc~Aa
(with the individual positions foreseen to be published in DR4) do not follow a
straight line, which renders the DR2 proper motions unreliable. Furthermore,
the periastron passage could be the cause of the unreliable astrometric
solution of \ktuc~Aa in Gaia DR3.

In constraining the orbit of \ktuc~B around A, \citet{tokovinin:2020} set the
orbital period to \qty{1200}{yr} to yield a total system mass of
\qty{2.22}{\Msun} that equals the sum of \qty{1.36}{\Msun} for \ktuc~Aa
\citep{fuhrmann:2017} and \qty{0.86}{\Msun} for \ktuc~B
\citep{pecaut:mamajek:2013}. Using our mass constraint for the combined system
of \ktuc~Aa and Ab and the semi-major axis of \qty{7.03(0.07)}{as} from
\citet{tokovinin:2020}, while neglecting any mass uncertainty of \ktuc~B, we
deduce the orbital period of \ktuc~B to be \qty{1122(28)}{yr}. We use the
inclination $i$ and position angle of ascending node $\Omega$ of the orbit of
\ktuc~B around A from \citet{tokovinin:2020} and those from \ktuc~Ab around Aa
from Table~\ref{tab_orbit} to compute the mutual inclination between these
orbits \citep{fekel:1981}. Due to the degeneracy of $\Omega$, the mutual
inclination is equally degenerate and can be either \qty{36.2(1.0)}{\degree}
or \qty{94.2(1.2)}{\degree}. Based on the stability criterion of
\citet[][eq.\ 13]{myllaeri:2018} with the parameter $A = \num{2.4}$ for
absolute stability and a strictness factor of $\lambda = \num{10}$, and the
unfavorable smaller mutual inclination of \qty{36.2}{\degree}, the current
configuration of the triple system \ktuc\ is stable for $\sim \qty{10}{Gyr}$,
which is longer than the remaining lifetime of \ktuc~Aa. Furthermore, tidal
circularization of the orbit of \ktuc~Ab is negligible due to the long
timescales \citep{goldreich:soter:1966, bodenheimer:2001}. However, the
distance between \ktuc~Aa and Ab at periastron is only $\approx \qty{0.3}{au}$.
Assuming the configuration is unchanged when \ktuc~Aa enters its late-stage
evolution, the expansion of \ktuc~Aa will alter the orbital dynamics.


\subsection{Previous Nondetections and Implications for Hot-exozodiacal Dust}
\label{subsect_nondetections}

The system of \ktuc~A was observed with PIONIER on 2012 July 24
\citep{ertel:2014, marion:2014}, 2013 August 10, and 2014 November 11
\citep{ertel:2016}. Only the data from 2012 were searched for signals from
stellar companions; the search region was restricted to a separation of
$< \qty{100}{mas}$ where clear companion signals are expected with PIONIER
\citep{absil:2011, marion:2014}. According to our orbit constraints, \ktuc~Ab
was located outside this search region for all PIONIER epochs (see green
points in Fig.~\ref{fig_orbit}).

For the 2012 and 2014 epochs, but not for 2013, a small deficit of the measured
squared visibility\footnote{PIONIER only measures the squared visibility and
not the visibility amplitude.} compared to the prediction from the stellar
photosphere alone was detected. With no companion detected, this deficit was
attributed to the presence of a hot exozodi with a flux ratio of
$\sim \qty{1}{\percent}$ compared to \ktuc~Aa, with the dust potentially being
of variable brightness due to the nondetection in 2013 \citep{ertel:2014,
ertel:2016}.

To assess whether \ktuc~Ab could have been the cause of the measured deficits,
we estimate the attenuation of its flux due to spatial filtering. The
transmission over the PIONIER field of view can be approximated by a Gaussian
with a full width at half maximum (FWHM) of \qty{400}{mas} assuming seeing of
\qty{0.8}{as} \citep{absil:2011}. For the smallest possible separation in 2014
of $\sim \qty{150}{mas}$ (see Fig.~\ref{fig_orbit}), about $\qty{70}{\percent}$
of the flux of \ktuc~Ab would have been transmitted. Assuming a flux ratio of
\ktuc~Ab to Aa in the $H$ band of \qty{1}{\percent} (estimated from the
$L$-band flux ratio and synthetic spectra, see
Sect.~\ref{subsect_brightness_spectral_type}), this would have been reduced by
spatial filtering to $\sim \qty{0.7}{\percent}$. The maximum deficit in squared
visibility that can be caused by a companion star is approximately 4 times the
flux ratio, thus $\Delta V^2 \lesssim \num{2.8e-2}$. Even considering that this
would be reduced by bandwidth smearing or worse transmission than assumed, it
is likely still similar to the deficit in squared visibility deduced by
\citet{ertel:2016} of $\Delta V^2 = \num{2.32(0.34)e-2}$ (twice the flux ratio
associated with the hot exozodi in their modeling, see their Fig.~1). Thus,
the companion could have been the origin of the measured deficit in squared
visibility and the hot-exozodi detection.

In 2013, the smallest possible separation was $\sim \qty{270}{mas}$ (see
Fig.~\ref{fig_orbit}), causing a transmission of $\sim \qty{30}{\percent}$.
In 2012, the separation was $\sim \qty{320}{mas}$, causing a transmission of
$\sim \qty{20}{\percent}$. The separation in both epochs was so large that the
fringe packets associated with the primary and the companion were not
overlapping for some of the baselines (ranging from $\sim \qty{9}{\m}$ to
\qty{31}{\m}) while they were still covered by the minimum scan length of
PIONIER \citep[][Sect.~3.2]{marion:2014}. This special case is not considered
by the PIONIER data reduction software \texttt{pndrs} \citep{le_bouquin:2011}.
A detailed assessment of this case would require an in-depth investigation of
the PIONIER data reduction process. Furthermore, a more accurate description
of how the electromagnetic wave front couples into the optical fiber
\citep[e.g.,][]{ruilier:cassaing:2001, wang:2021} under the actual atmospheric
conditions compared to the Gaussian approximation by \citet{absil:2011} is
needed. Both are beyond the scope of this article. While it appears unlikely
to us that \ktuc~Ab could have caused the entire deficit in squared visibility
of $\Delta V^2 = \num{2.86(0.34)e-2}$ measured in 2012 \citep{ertel:2014}, we
cannot judge by how much \ktuc~Ab influenced the hot-exozodi detection in 2012.
If the 2012 detection was caused by \ktuc~Ab, the nondetection in 2013 could
be explained by an unfavorable sampling of the spatial frequencies or the
baseline orientation projected onto the sky.

For the MATISSE observations on 2019 July 9 and 11, \ktuc~Ab cannot explain the
deep deficit of the visibility amplitude measured by
\citet{kirchschlager:2020}. The visibility amplitudes (of our new reductions)
show that the deficit compared to the expected signal from \ktuc~Aa is nearly
constant over spatial frequency, consistent with an over-resolved brightness
distribution \citep[e.g.,][]{di_folco:2007, ertel:2014} as modeled by
\citet{kirchschlager:2020}. Furthermore, the visibility amplitudes for the
shorter $\sim \qty{35}{\meter}$ baselines (not affected by bandwidth smearing)
at this lower overall visibility level appear modulated by an oscillating
signal, in agreement with an off-axis point source with a flux ratio of
$\sim \qty{0.5}{\percent}$ to the primary \citep[a companion with flux ratio
$f$ causes a modulation in visibility amplitudes of $\Delta |V| \leq 2f$,
e.g.,][]{tsishchankava:2025}. This agrees well with our estimate for the
attenuated flux ratio based on the closure phases
(Table~\ref{tab_astrometry}).

We conclude that, while \ktuc~Ab left imprints in the visibility amplitudes
analyzed by \citet{kirchschlager:2020}, it cannot account for the overall
measured deficit in the visibility amplitude that has been attributed to a hot
exozodi.


\subsection{Implications for the Planetary System}
\label{subsect_planetary_system}

The coexistence of \ktuc~Ab with a hot exozodi motivates dynamical studies of
how the binary star interacts with the planetary system. The periastron of
\ktuc~Ab coincides with the constrained location of the hot exozodi
\citep{kirchschlager:2020} and generally where hot exozodis are found
\citep{absil:2006, akeson:2009, defrere:2011, lebreton:2013,
kirchschlager:2017, stuber:2023b, ollmann:2025}, which will have a strong
influence on the dust dynamics and lifetime during periastron passage: For
instance, the added emission of \ktuc~Ab will reduce sublimation timescales
\citep[e.g.,][]{kobayashi:2009, lebreton:2013, sezestre:2019, pearce:2020} and
its gravitational effect might disrupt the orbits of dust grains and reduce the
efficiency of any potential dust-trapping mechanism around the primary
\ktuc~Aa, such as gas drag \citep{lebreton:2013, pearce:2020}, magnetic field
lines \citep{czechowski:mann:2010, su:2013, rieke_GH:2016, stamm:2019,
kimura:2020}, or the differential Doppler effect \citep{burns:1979, rusk:1987,
sezestre:2019}. Furthermore, \ktuc~Ab might trap dust in its own vicinity.

However, the periastron passage is rapid, and \ktuc~Ab spends most of the time
at larger distances of several astronomical units from the primary. If the hot
dust is short-lived and continuously supplied, for instance by exocomets, the
hot exozodi would quickly recover from any impact by the periastron passage.
While cometary supply as the sole cause of hot exozodis is disfavored, because
it requires unphysically large comet-inflow rates \citep{pearce:2022b}, this
periodic dust-level effect could still occur if some efficient trapping
mechanism were also present.

Whereas the large apoastron of \ktuc~Ab allows for a change in the dominant
physical processes in the primary's close circumstellar environment, \ktuc~Ab
might interact with a possible cold debris disk \citep[e.g.,][]{matthews:2014,
hughes:2018, wyatt:2020} at tens of astronomical units. While no far-infrared
excess of the \ktuc~A system has been found \citep{gaspar:2013,
sibthorpe:2018}, which agrees with the statistical lack of far-infrared
detected debris disks in binary systems \citep{rodriguez:2012, rodriguez:2015,
yelverton:2019}, the existence of the hot exozodi gives rise to the possibility
of a dust reservoir formed by a planetesimal population further out than the
apoastron of \ktuc~Ab; at those distances also the gravitational influence of
the wider companion \ktuc~B might have to be considered. The gravitational
influence of both \ktuc~Ab and B would shape the planetesimal distribution
\citep[e.g.,][]{faramaz:2014, pearce:2014, regaly:2018, sende:loehne:2019,
stuber:2023a} and provide means to scatter planetesimals onto orbits that
graze the primary \citep{faramaz:2015, faramaz:2017} where they could replenish
the hot-dust distribution. Even a planetesimal distribution that is penetrated
by the orbit of \ktuc~Ab cannot be excluded \citep{pearce:2021}.

Lastly, while the radiation pressure (see \citeauthor{burns:1979},
\citeyear{burns:1979}; \citeauthor{gustafson:1994}, \citeyear{gustafson:1994}
for reviews on forces acting on dust grains) exerted by the emission of
M dwarfs is not effective in repelling dust grains, their stellar winds can be
effective in repelling or dragging dust grains \citep[e.g.,][]{plavchan:2005,
plavchan:2009, augereau:beust:2006, strubbe:chiang:2006, reidemeister:2011,
schueppler:2014, schueppler:2015, matthews:2015}, which affects the interplay
of transport and collisional processes in the entire system.


\section{Summary}
\label{sect_summary}

We detected a low-mass stellar companion to \ktuc~Aa with infrared
interferometry using the VLTI instruments MATISSE and GRAVITY. With an
$L$-band contrast of \qty{1.32}{\percent}, this is the highest-contrast
stellar companion ever detected with MATISSE. Furthermore, we demonstrated
that MATISSE is able to deliver relative astrometry with milliarcsecond
precision in imaging mode, which we validated with GRAVITY. Measuring precise
flux ratios for such binaries with contrasts of $\sim \num{1e-2}$ and
separations of $\gtrsim \qty{100}{mas}$ is challenging with single observations
for both MATISSE and GRAVITY, probably due to imperfect centering of the source
and atmospheric conditions; multiple observations are advised to statistically
mitigate this issue.

The new companion \ktuc~Ab is an M dwarf of spectral type M3.5$\,$V to
M4.5$\,$V that moves around \ktuc~Aa with a period of $\approx \qty{8.14}{yr}$
on a highly eccentric orbit with $e \approx 0.94$ and a semi-major axis of
$\approx \qty{4.8}{au}$. While \ktuc~Ab might have caused the putative
detection of hot-exozodiacal dust around \ktuc~Aa in 2012 and 2014, it cannot
negate the hot-dust detection in 2019.

This coexistence of hot dust and the stellar companion motivates dynamical
studies of this intriguing planetary system, governing, for instance, how
\ktuc~Ab interacts with the hot-dust distribution during its periastron passage
or how it might excite unseen planetesimals onto cometary orbits that can
replenish the dust in situ.


\section*{Acknowledgments}
Special thanks go to Alexis Matter for his invaluable assistance regarding
every aspect of MATISSE. Further thanks go to Walter Jaffe for his aid in
describing the spatial filtering of MATISSE.
Observations were collected at the European Organisation
for Astronomical Research in the Southern Hemisphere under ESO programs
0103.C-0725, 109.23HL, 110.24AJ, 111.24S6, 114.27F3, 114.28HE, and 114.28HE.
This work is supported by the National Aeronautics and Space Administration
(NASA) through grants 80NSSC23K1473 (T.A.S., W.C.D., V.F.-G., J.P.S., S.E.),
80NSSC21K0394 (V.F.-G., S.E.), 80NSSC23K0288 (T.A.S., V.F.-G., S.E.), and
80NSSC21K0593 (G.W.);
by the European Research Council (ERC) under the European Union’s Horizon
2020 research and innovation program through programs ERC-2019-StG-851622
(F.K.) and ERC-2019-CoG-866070 (G.G. and D.D.);
by the DFG through grants WO 857/15-2 (S.W., K.O) and KR 2164/15-2 (A.V.K.);
by a UKRI Stephen Hawking Fellowship and a Warwick Prize Fellowship, the
latter made possible by a generous philanthropic donation (T.D.P.);
by the Programme National de Planétologie (PNP) of CNRS-INSU in France,
through the EPOPEE (Etude des POussières Planétaires Et Exoplanétaires)
project (J.-C.A., P.P.);
by the Research Foundation - Flanders (FWO) under grant 1234224N (R.L.);
by the French Agence Nationale de la Recherche (ANR) through grant
ANR-21-CE31-0011 (R.G.P.);
by a state scholarship awarded by Kiel University (K.T.).
O.A. is a Senior Research Associate of the Fonds de la Recherche
Scientifique – FNRS.
This work has made use of data from the European Space Agency (ESA) mission
{\it Gaia}\footnote{\url{https://www.cosmos.esa.int/gaia}}, processed by the
{\it Gaia} Data Processing and Analysis Consortium
(DPAC)\footnote{\url{https://www.cosmos.esa.int/web/gaia/dpac/consortium}}
with funding provided by national institutions, in particular the institutions
participating in the {\it Gaia} Multilateral Agreement;
of data from the Two Micron All Sky Survey,
which is a joint project of the University of Massachusetts and the Infrared
Processing and Analysis Center/California Institute of Technology, funded by
NASA and the National Science Foundation;
of collaborations and/or information exchange within NASA’s Nexus for Exoplanet
System Science (NExSS) research coordination network sponsored by NASA’s
Science Mission Directorate;
of the Jean-Marie Mariotti Center (JMMC) Optical Interferometry Database and
the JMMC services \texttt{Aspro}, \texttt{OIFits Explorer}, \texttt{SearchCal},
which involves the JSDC and JMDC catalogs, and the JMMC Expertise Center for
the User Support\footnote{\url{https://www.jmmc.fr/suv}}.
This work used High Performance Computing resources supported by the University
of Arizona TRIF, UITS, and Research, Innovation, and Impact (RII) and maintained
by the University of Arizona Research Technologies department.


\facility{VLTI (GRAVITY, MATISSE, NAOMI)}


\software{
    \texttt{Aspro} \citep{bourges:2016},
    \texttt{astropy} \citep{astropy_collaboration:2013,
    astropy_collaboration:2018, astropy_collaboration:2022},
    \texttt{emcee} \citep{foreman-mackey:2013},
    \texttt{Ipython} \citep{perez:granger:2007},
    GRAVITY data reduction software \citep{lapeyrere:2014},
    \texttt{Jupyter} notebooks \citep{kluyver:2016},
    \texttt{KDEpy} (\url{https://kdepy.readthedocs.io}),
    MATISSE data reduction software \citep{millour:2016},
    \texttt{Matplotlib} \citep{hunter:2007},
    \texttt{NumPy} \citep{harris:2020},
    \texttt{OIBD} (\url{http://oidb.jmmc.fr}),
    \texttt{OIFITS} \citep{duvert:2017},
    \texttt{OIFits Explorer}
    (\url{https://www.jmmc.fr/english/tools/data-analysis/oifits-explorer/}),
    \texttt{orbitize!}\ \citep{blunt:2020},
    \texttt{PMOIRED} \citep{merand:2022, merand:2024},
    \texttt{ptemcee} \citep{vousden:2016, vousden:2021},
    \texttt{python} (\url{https://www.python.org/}),
    \texttt{SearchCal}
    (\url{https://www.jmmc.fr/english/tools/proposal-preparation/search-cal/}),
    \texttt{SIMBAD} Astronomical Database \citep{wenger:2000},
    \texttt{VizieR} \citep{ochsenbein:2000}.
}


\appendix


\section{Treatment of Bandwidth Smearing by PMOIRED}
\label{appdx_sect_bandwidth_smearing}

Modeling of interferometric data often treats light as monochromatic. However,
for a finite resolving power $R = \lambda/\delta\lambda$, the spectral
resolution element $\delta\lambda$ (also called spectral bandwidth or spectral
channel) has a finite length. In this case, bandwidth smearing occurs if the
angular separation $\rho$ between two sources along the projected baseline
becomes comparable to the width of a fringe packet associated with a single
point source, thus $\rho \gtrsim R\lambda /B$ (see, e.g.,
\citeauthor{lachaume:berger:2013} \citeyear{lachaume:berger:2013} for a
detailed description). For our observations, this is relevant only for MATISSE
observations in LOW-LM spectral resolution.

In case of a binary star with fixed separation, the fringe visibility and phase
are modulated with increasing spatial frequencies $B/\lambda$. PMOIRED models
bandwidth smearing by convolving the simulated observables in the spectral
domain with a Gaussian. Its FWHM equals the spectral resolution element, which
is specified by its oversampling by detector pixels (set by the \texttt{wl
kernel} keyword in \texttt{setupFit} of \texttt{PMOIRED}); for the LM-band arm
of MATISSE, the oversampling is $\approx 5$. If smearing occurs over several
detector pixels, the observables have to be computed with higher spectral
resolution (set by the \texttt{smear} key word) before convolution, and then
reduced to the correct resolution afterwards. This effect was not considered
for our case, because the high oversampling by MATISSE already causes the
observables to be computed at a sufficiently high resolution.

If bandwidth smearing is not accounted for, the binary flux ratio can be
underestimated. Furthermore, for different baseline lengths, the modulation
amplitude gets damped by different degrees, which biases a joint analysis of
multiple baselines. For our largest separation (2019 July,
$\rho \sim \qty{300}{mas}$), accounting for bandwidth smearing with
\texttt{PMOIRED} increases the determined flux ratio from
\qty{0.13(0.01)}{\percent} to \qty{0.51(0.05)}{\percent}. Our treatment is
validated by the remaining trend of binary flux ratio with separation that
agrees with the expected behavior caused by spatial filtering inside MATISSE
(Appendix.~\ref{appdx_sect_matisse_attenuation}).


\section{MCMC Setup}
\label{appdx_sect_MCMC}

For our MCMC setup, we use default \texttt{orbitize!}\ priors for the following
orbital parameters: uniform in \prior{0}{1} for the eccentricity $e$ and
dimensionless epoch of periastron passage $\tau$, uniform in \prior{0}{2\pi}
for the argument of periastron $\omega$, log-uniform in
\prior{\qty{e-3}{au}}{\qty{e4}{au}} for the semi-major axis $a$, and $\sin{i}$
for the inclination $i$ with $i$ uniform in \prior{0}{\pi}. For the position

angle of ascending node $\Omega$ we choose a prior uniform in \prior{0}{\pi} to
restrict the orbits to one of the degenerate modes and aid convergence. For the
parallax of the system, we apply a Gaussian prior with a mean of
\qty{47.65}{mas} and a standard deviation of \qty{0.02}{mas}. For the total
system mass of primary and companion $M\sbs{tot}$ we apply a prior uniform in
\prior{\qty{1}{\Msun}}{\qty{2.5}{\Msun}}; this interval ranges from the lower
mass end of F-type stars up to the upper mass end of a pair of F- and M-type
stars. As verified by the resulting posterior, this allowed mass interval well
encloses the constrained mass and hence is not overly restrictive. Furthermore,
as the coverage of the orbital phase is $> \qty{40}{\percent}$, the use of
Campbell orbital elements and uniform priors does not bias our results
\citep{lucy:2014, oneil:2019}.

\begin{figure*}
    \centering
    \includegraphics[width=0.93\textwidth]{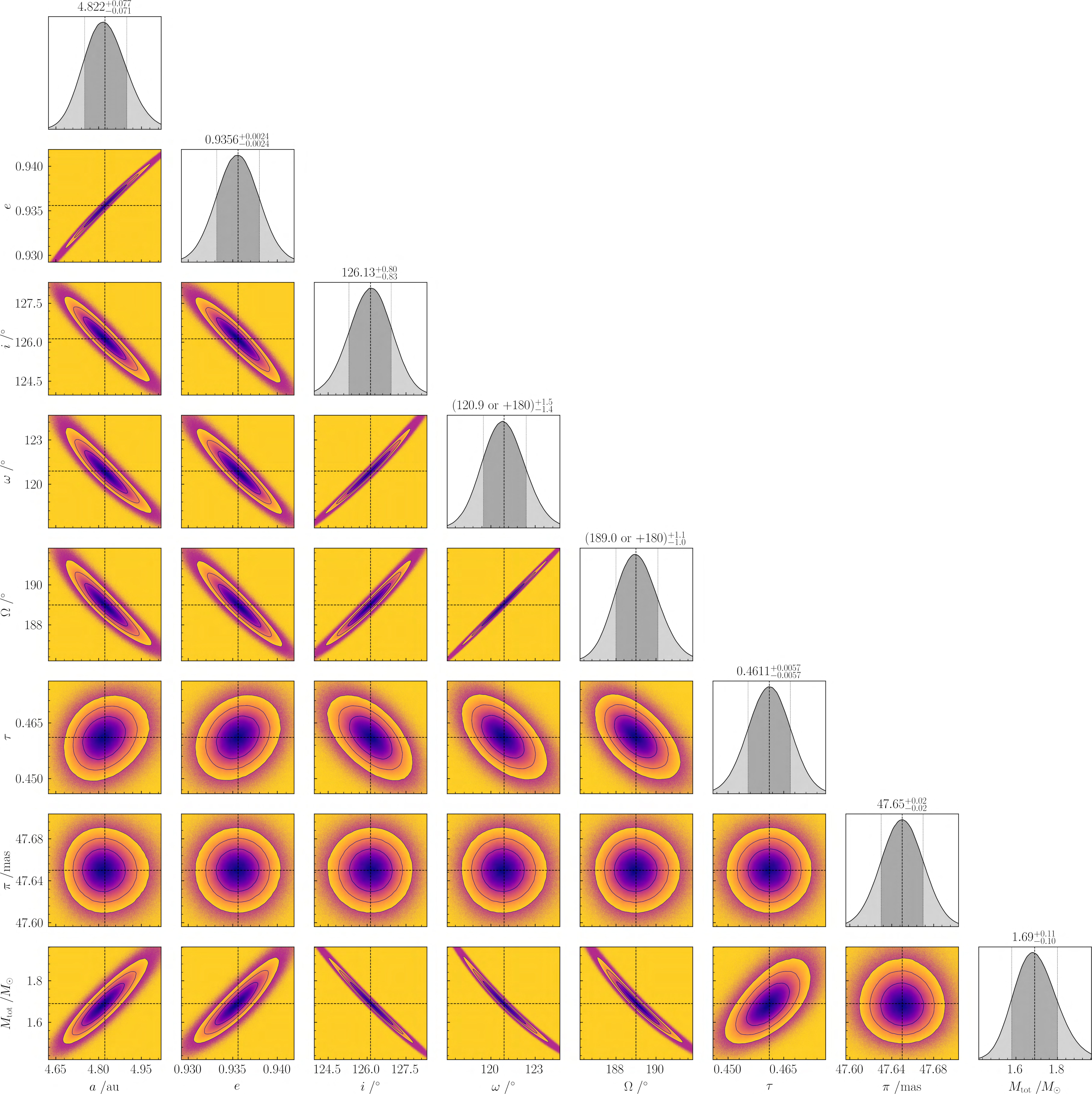}
    \caption{
        Corner plot showing the converged \texttt{orbitize!}\ posterior sample
        thinned by a factor of 10 (i.e., $10^7$ samples). We show gray 200-bin
        histograms for each orbital element along the diagonal with a KDE
        overplotted in black. Dashed and dotted lines mark the median and
        0.16/0.84 quantiles computed from the complete posterior sample. The
        solid contours in the off-diagonal panels show the \num{2}$\sigma$,
        \num{1.5}$\sigma$, and \num{1}$\sigma$ equivalent levels estimated from
        two-dimensional KDEs, containing $\approx \qty{86.5}{\percent}$,
        $\approx \qty{67.5}{\percent}$, and $\approx \qty{39.3}{\percent}$ of
        the samples, with $\sigma$ being the standard deviation of a circular
        bivariate Gaussian. Samples are scattered below the 2$\sigma$
        equivalent levels, and the two-dimensional KDEs are shown with filled
        contours above this level.
    }
    \label{fig_corner_plot}
\end{figure*}
\begin{figure*}
    \centering
    \includegraphics{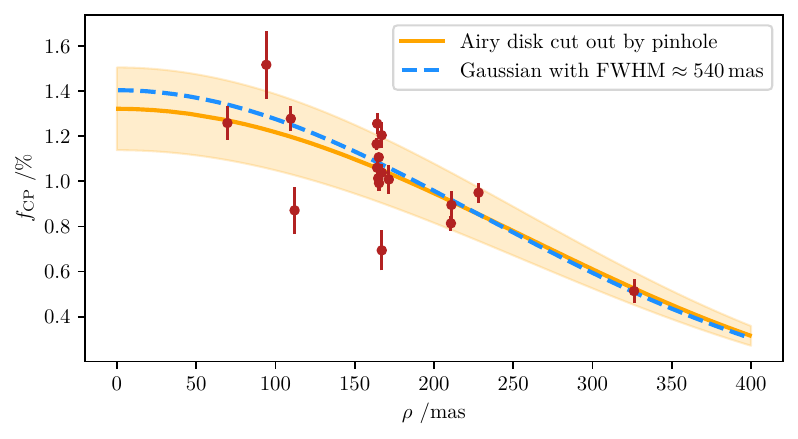}
    \caption{
        Attenuated flux ratios $f\sbs{CP}$ (red points) as measured directly
        from the interferometric closure phases, compared to the prediction
        assuming transmission loss by spatially filtering an Airy disk
        separated from the pinhole center by $\rho$ (solid orange line). The
        nonattenuated flux ratio for $\rho = \qty{0}{mas}$ is computed from the
        mean of the flux ratios corrected from attenuation; the faint orange
        region covers their standard deviation. This treatment is consistent
        with a Gaussian description of the attenuation (dashed blue line).
    }
    \label{fig_spatial_attenuation}
\end{figure*}

We employ \num{1000} walkers, each using \num{20} temperatures (the
\texttt{orbitize!}\ default). Walkers at higher temperatures explore a
flattened (\textit{tempered}) posterior distribution that can more easily be
sampled. Periodically, there are swaps in the position of walkers of adjacent
temperatures, promoting the exploration of the entire parameter space, also for
low-temperature walkers. Only the samples of the lowest-temperature walkers
that explore the unaltered posterior distribution are analyzed. 

First, we perform a blind run with walker positions randomly initialized within
the prior constraints and perform a total of \num{5e8} steps. We apply a
\textit{thinning} of \num{100}, that is, only every 100th step is saved. Thus,
the chain length of each of the 1000 lowest-temperature walkers for each
orbital parameter equals \num{5000}. The chains appear visually converged after
$\sim \num{500}$ steps. For each parameter, we estimate from the chains the
integrated autocorrelation time $t$, that is, the number of adjacent steps in
the chain that are correlated. The autocorrelation times range from
$t \approx \num{2.5}$ for the parallax up to $t \approx \num{39}$ for the
eccentricity. As our chains are much longer than $50t$, our estimates of $t$
are reliable \citep{foreman-mackey:2021}, and the final walker positions are
independent from the start positions. We conclude that the chains are fully
converged.

From the final walker positions of this first run, we start a second run with
\num{e8} steps, without thinning, to sample the posterior distribution for
analysis. The individual chain length is \num{e5}, and autocorrelation times
are $t \approx \num{8}$ for the parallax and $t \approx \num{6}$ for all
other parameters. Figure~\ref{fig_corner_plot} shows the posterior distribution
(thinned by a factor of \num{10} for computational reasons), computed following
the procedure of \citet{weible:2025} and using kernel density estimates (KDE)
computed with \texttt{KDEpy.FFTKDE},\footnote{
\href{https://kdepy.readthedocs.io/en/latest/API.html\#fftkde}{
https://kdepy.readthedocs.io/en/latest/API.html\#fftkde}} which is based on
the \texttt{statsmodels}\footnote{
\href{https://www.statsmodels.org/stable/dev/index.html}{
https://www.statsmodels.org/stable/dev/index.html}} implementation
\citep{seabold:perktold:2010}.


\section{Spatial Filtering and Attenuation of Off-axis Sources for MATISSE}
\label{appdx_sect_matisse_attenuation}

MATISSE uses pinholes at an intermediate focus to spatially filter the incoming
beam of each telescope. The pinholes' angular diameter for the LM-band arm is
$\num{1.5}\lambda/D$ in the pupil plane. If the point-spread function of a
source is not centered on the pinhole, for instance, in the case of a companion
separated from the centered primary star, less light is transmitted through the
pinhole compared to the case of the fully centered point-spread function. We
estimate the transmission $T$ for an off-axis source by computing the
transmitted flux of an Airy disk through the pinhole, divided by the
transmitted flux of an Airy disk centered on the pinhole. Then we correct the
flux ratio affected by spatial attenuation $f\sbs{CP}$ via $f = f\sbs{CP}/T$
(Table~\ref{tab_observations}) and compute the final flux ratio in the $L$
band as the mean of all measurements with its standard error
$f_L \approx \qty{1.32(0.04)}{\percent}$; the standard deviation of the
measurement sample is $\approx \qty{0.18}{\percent}$. Assuming $f_L$ as the
nonattenuated flux ratio for a point-spread function centered on the pinhole,
we compare the predicted attenuated flux ratios with our measurements in
Fig.~\ref{fig_spatial_attenuation} and find good agreement. Furthermore, this
is consistent with a spatial attenuation following a Gaussian profile with a
FWHM of $\approx \qty{540}{mas}$ as obtained from a least-squares fit to the
measurements. Thus, this can be used as an approximation of the flux
attenuation by spatial filtering. However, this treatment is only possible if
attenuation by spectral-bandwidth smearing is either negligible or has been
accounted for in the interferometric analysis (see
Sect.~\ref{appdx_sect_bandwidth_smearing}). A reason for the remaining scatter in
Fig.~\ref{fig_spatial_attenuation} could be inaccuracies in centering the Airy
disk in the pinhole.


\section{Determination of the $K$\sbs{s}-band Magnitude of the Binary \ktuc~A}
\label{appdx_sect_magnitude}

To determine the $K$\sbs{s}-band magnitude of the binary \ktuc~A,
$m_\textrm{$K$\sbs{s},$\,$A}$, from the joint magnitude of \ktuc~A and B,
$m_\textrm{$K$\sbs{s},$\,$AB} = \num{3.876}$ \citep{cutri:2003}, we first
estimate the $K$\sbs{s}-band magnitude of \ktuc~B,
$m_\textrm{$K$\sbs{s},$\,$B}$. With its $V$-band magnitude of
$m_\textrm{$V$,$\,$B} = \num{7.58}$ and spectral type of K1$\,$V
\citep{corbally:1984}, we obtain the color from \citet{pecaut:mamajek:2013} as
$m_V - m_\textrm{$K$\sbs{s}} = 2.06$\footnote{From the updated online version
of their catalog in version 2022.04.16,
\href{https://www.pas.rochester.edu/~emamajek/EEM_dwarf_UBVIJHK_colors_Teff.txt}{
https://www.pas.rochester.edu/$\sim$emamajek/EEM\_\\dwarf\_UBVIJHK\_colors\_Teff.txt}.}
which results in $m_\textrm{$K$\sbs{s},$\,$B} = \num{5.52}$. Using standard
magnitude relations, we compute the magnitude of \ktuc~A from
$m_\textrm{$K$\sbs{s},$\,$AB}$ and $m_\textrm{$K$\sbs{s},$\,$B}$ to be
$m_\textrm{$K$\sbs{s},$\,$A} = \num{4.15}$.


\section*{ORCID iDs}

{\small
\noindent T. A. Stuber \orcidlink{0000-0003-2185-0525}%
\href{https://orcid.org/0000-0003-2185-0525}{https://orcid.org/0000-0003-2185-0525}\\
A. Mérand \orcidlink{0000-0003-2125-0183}%
\href{ttps://orcid.org/0000-0003-2125-0183}{https://orcid.org/0000-0003-2125-0183}\\
F. Kirchschlager \orcidlink{0000-0002-3036-0184}%
\href{https://orcid.org/0000-0002-3036-0184}{https://orcid.org/0000-0002-3036-0184}\\
S. Wolf \orcidlink{0000-0001-7841-3452}%
\href{https://orcid.org/0000-0001-7841-3452}{https://orcid.org/0000-0001-7841-3452}\\
G. Weible \orcidlink{0000-0001-8009-8383}%
\href{https://orcid.org/0000-0001-8009-8383}{https://orcid.org/0000-0001-8009-8383}\\
O. Absil \orcidlink{0000-0002-4006-6237}%
\href{https://orcid.org/0000-0002-4006-6237}{https://orcid.org/0000-0002-4006-6237}\\
T. D. Pearce \orcidlink{0000-0001-5653-5635}%
\href{https://orcid.org/0000-0001-5653-5635}{https://orcid.org/0000-0001-5653-5635}\\
G. Garreau \orcidlink{0000-0001-6282-1339}%
\href{https://orcid.org/0000-0001-6282-1339}{https://orcid.org/0000-0001-6282-1339}\\
J.-C. Augereau \orcidlink{0000-0002-2725-6415}%
\href{https://orcid.org/0000-0002-2725-6415}{https://orcid.org/0000-0002-2725-6415}\\
W. C. Danchi \orcidlink{0000-0002-9209-5830}%
\href{https://orcid.org/0000-0002-9209-5830}{https://orcid.org/0000-0002-9209-5830}\\
D. Defrère \orcidlink{0000-0003-3499-2506}%
\href{https://orcid.org/0000-0003-3499-2506}{https://orcid.org/0000-0003-3499-2506}\\
\mbox{V. Faramaz-Gorka \orcidlink{0000-0001-6403-841X}%
\href{https://orcid.org/0000-0001-6403-841X}{https://orcid.org/0000-0001-6403-841X}}\\
J. W. Isbell \orcidlink{0000-0002-1272-6322}%
\href{https://orcid.org/0000-0002-1272-6322}{https://orcid.org/0000-0002-1272-6322}\\
J. Kobus \orcidlink{0000-0002-3741-5950}%
\href{https://orcid.org/0000-0002-3741-5950}{https://orcid.org/0000-0002-3741-5950}\\
A. V. Krivov \orcidlink{0009-0009-4573-2612}%
\href{https://orcid.org/0009-0009-4573-2612}{https://orcid.org/0009-0009-4573-2612}\\
R. Laugier \orcidlink{0000-0002-2215-9413}%
\href{https://orcid.org/0000-0002-2215-9413}{https://orcid.org/0000-0002-2215-9413}\\
K. Ollmann \orcidlink{0009-0003-6954-5252}%
\href{https://orcid.org/0009-0003-6954-5252}{https://orcid.org/0009-0003-6954-5252}\\
R. G. Petrov \orcidlink{0000-0003-4759-6051}%
\href{https://orcid.org/0000-0003-4759-6051}{https://orcid.org/0000-0003-4759-6051}\\
P. Priolet \orcidlink{0009-0000-4227-5449}%
\href{https://orcid.org/0009-0000-4227-5449}{https://orcid.org/0009-0000-4227-5449}\\
J. P. Scott \orcidlink{0009-0000-3882-9242}%
\href{https://orcid.org/0009-0000-3882-9242}{https://orcid.org/0009-0000-3882-9242}\\
K. Tsishchankava \orcidlink{0009-0002-9371-0740}%
\href{https://orcid.org/0009-0002-9371-0740}{https://orcid.org/0009-0002-9371-0740}\\
S. Ertel \orcidlink{0000-0002-2314-7289}%
\href{https://orcid.org/0000-0002-2314-7289}{https://orcid.org/0000-0002-2314-7289}
}


\bibliography{bibliography}{}
\bibliographystyle{aasjournal}


\end{document}